\documentclass[aaspp4,11pt,preprint,apjfonts]{emulateapj}

\newcommand{\um}{$\mu$m}

\newcommand{\Msol}{M$_{\odot}$}
\newcommand{\Zsol}{Z$_{\odot}$}
\newcommand{\HI}{H\,{\sc i}}
\newcommand{\HII}{H\,{\sc ii}}
\newcommand{\HeI}{He\,{\sc i}}
\newcommand{\HeII}{He\,{\sc ii}}
\newcommand{\ArIII}{[Ar\,{\sc iii}]}
\newcommand{\SIV}{[S\,{\sc iv}]}
\newcommand{\NeII}{[Ne\,{\sc ii}]}
\newcommand{\NeIII}{[Ne\,{\sc iii}]}
\newcommand{\SIII}{[S\,{\sc iii}]}

\newcommand{\OIV}{[O\,{\sc iv}]}
\newcommand{\FeII}{[Fe\,{\sc ii}]}
\newcommand{\FeIII}{[Fe\,{\sc iii}]}
\newcommand{\SiII}{[Si\,{\sc ii}]}

\shorttitle{Mid-IR diagnostics}
\shortauthors{Snijders et al.}

\begin{document}

\title{Mid-infrared diagnostics of starburst galaxies: clumpy, dense structures in star-forming regions in the Antennae (NGC 4038/4039)\altaffilmark{1}}

\author{Leonie Snijders\altaffilmark{2},
Lisa J. Kewley\altaffilmark{3} \&
Paul P. van der Werf\altaffilmark{2}}

\altaffiltext{1}{Based on observations collected at the European Southern Observatory, Paranal, Chile, under program no. 075.B-0727(A), 075.B-0791(A), and 
69.B-0688(A)} 
\altaffiltext{2}{Leiden Observatory, Leiden University, PO Box 9513, 2300 RA Leiden, The Netherlands}
\altaffiltext{3}{University of Hawaii, 2680 Woodlawn Drive, Honolulu, USA}

\begin{abstract}
Recently, mid-infrared instruments have become available on several
large ground-based telescopes, resulting in data sets with
unprecedented spatial resolution at these long wavelengths. In this
paper we examine 'ground-based-only' diagnostics, which can be used in
the study of star-forming regions in starburst galaxies. By combining
output from the stellar population synthesis code {\it Starburst 99}
with the photoionization code {\it Mappings}, we model stellar
clusters and their surrounding interstellar medium, focusing on the
evolution of emission lines in the N- and Q-band atmospheric windows
(8 -- 13 and 16.5 -- 24.5 \um ~respectively) and those in the
near-infrared. We address the detailed sensitivity of various emission
line diagnostics to stellar population age, metallicity, nebular
density, and ionization parameter. Using our model results, we analyze
observations of two stellar clusters in the overlap region of the
Antennae galaxies obtained with VLT Imager and Spectrometer for mid
Infrared (VISIR). We find evidence for clumpy, high density, ionized
gas. The two clusters are young (younger than 2.5 and 3 Myr
respectively), the surrounding interstellar matter is dense ($\ge$
10$^4$ cm$^{-3}$) and can be characterized by a high ionization
parameter (logU $\ge$ -1.53). Detailed analysis of the mid-infrared
spectral features shows that a (near-)homogeneous medium cannot
account for the observations, and that complex structure on scales
below the resolution limit, containing several young stellar clusters
embedded in clumpy gas, is more likely.

\end{abstract}


\section{Introduction}
With a star formation rate (SFR) of several tens to several hundreds
of solar masses a year, starburst galaxies form stars at a much higher
rate than regular gas-rich galaxies. As a result, starburst galaxies
empty their gas reservoirs quickly during a short, vigorous burst of
star-formation. This short-lived starburst phase is generally
initiated by interactions with neighbor galaxies, causing distortions
in the galaxy's gravitational potential. As a consequence molecular
gas piles up, and in the resulting concentrations of very dense gas
active star formation takes place. Although locally rare, because of
the limited duration of the starburst phase and the relatively low
galaxy pair density in the local universe, starburst galaxies are a
very important piece in the puzzle of galaxy evolution. In a very
short time the galaxy's morphology and stellar content are
significantly altered, and the interstellar and even intergalactic
medium is rapidly enriched. At higher redshifts starburst galaxies are
a much more common phenomenon, and sub-mm observations indicate that
the intense outbursts of star formation in these objects might be the
dominant mode of star formation in the high--z universe
\citep[e.g.][and many others]{Heckman:98, Blain:99, Smail:02}. Despite numerous
studies of starburst galaxies, several issues remain unsolved, like
the exact mechanism that ignites and shuts off the starburst phase,
and the mass function of the individual stellar populations formed
during the bursts.

In this work we focus on the individual star-forming regions formed
during these starbursts. With a wealth of young, bright stellar
clusters, starburst galaxies are excellent objects for the study of
the physical processes of star formation. This is why starburst
galaxies play an important role in the debate on the universality of
the stellar Initial Mass Function \citep[IMF,
][]{Kroupa:07}. Since star formation takes place in dusty giant
molecular clouds, the active star-forming regions in starburst
galaxies generally suffer from high obscuration. This makes the
(mid-)infrared a favorable regime for the study of the youngest, most
massive, deeply embedded stellar populations. Analysis of mid-infrared
nebular emission lines of clusters in starburst galaxies have led to
various different conclusions on the IMF of these objects. Contrary to
what one would expect for young, massive stellar populations, ISO
measured relatively low values for the \NeIII/\NeII ~ratio, indicating
a soft UV radiation field. \cite{Thornley:00} suggest that the soft
radiation field is a result from the aging of stellar populations with
a 'regular' IMF (up to M$_{\rm up}$ $\sim$ 50 -- 100 \Msol), formed in
short-lived bursts of star formation (1 -- 10 Myr). Alternative
explanations are a truncated IMF, lacking massive stars, or massive
stars spending a considerable fraction of their lifetime embedded in an
ultra compact (UC)\HII ~region, hidden from view even at mid-infrared
wavelengths \citep[][]{Rigby:04}. In relation to this discussion,
\citet{Martin:02} already pointed out the need for a more
comprehensive study of mid-infrared diagnostics, since the neon ratio
is not only sensitive to radiation hardness, but to metallicity,
nebular density and ionization parameter as well.

Over the last decades the field of mid-infrared astronomy received an
enormous impulse from observations with space telescopes; from
Infrared Astronomical Satellite (IRAS) in the eighties, Infrared Space
Observatory (ISO) in the nineties and currently from the Spitzer Space
Telescope. However, as discussed by \cite{Martin:05}, the
interpretation of large aperture mid-infrared observations obtained
with space telescopes can be challenging, especially for extragalactic
objects. Ground-based mid-infrared astronomy faces additional
challenges due to the characteristics of the earth's atmosphere, but
thanks to rapid progress in detector technology, several mid-infrared
instruments have become available on large ground-based observatories
in the past five years. Compared to observations from space,
ground-based work offers the considerable advantage of a more detailed
view of the objects of interest, due to the much higher spatial
resolution that can be obtained. For example, the typical size of a
star-forming region is a few to a few tens of parsecs. With its 85 cm
mirror, Spitzer can resolve details on a 50 parsec scale in objects
out to 4 Mpc distance. With an 8 meter class telescope similar scales
can be resolved out to 40 Mpc, increasing the volume that can be
probed at this resolution by a factor of a thousand. Unfortunately, the
atmosphere is opaque to most infrared radiation, making it impossible
to access the full infrared wavelength regime from the ground. This
means that we are limited to a number of specific atmospheric windows
of (reasonably) good transmission; the L band between 3.0 and 4.0 \um,
the M band in the range of 4.6 to 5.2 \um, the N band from 8 to 13
\um~ and the Q band between 16.5 and 24.5 \um. The aim of this paper
is to explore the spectral diagnostic features available for the study
of star-forming regions within these windows, combining near- and
mid-infrared wavelengths (J, H, K, L, N and Q band, roughly covering 1
-- 25 \um). We will address the behavior of near- and mid-infrared
emission line ratios as a function of cluster age, gas pressure,
metallicity, and ionization parameter.

In Section 2 we discuss the spectral features observable with
ground-based facilities and the codes used in our model efforts, the
stellar population synthesis model, {\it Starburst 99}, and the
photoionization code {\it Mappings}, as well as the model grid input
parameters. The model results are presented in Section 3, and compared
with existing models in Section 4. In Section 5 diagnostic diagrams
are introduced, which are compared to mid-infrared data of \HII
~regions from the literature in Section 6. In Section 7 we present
mid-infrared spectroscopic observations of clusters in the Antennae
overlap regions. In Section 8 these data are analyzed and our results
are discussed. Finally, we summarize our findings in Section 9.

\section{Modeling the emission line spectra}

Our work focuses on stellar populations created during vigorous bursts
of star formation in starburst galaxies. These stellar populations are
young, dust-enshrouded star clusters, still embedded in their natal
clouds. To model these systems we combine the stellar population
synthesis model, {\it Starburst 99} \citep{Leitherer:99}, with the
photoionization code {\it Mappings IIIr} \citep{Dopita:00,
Dopita:02,Groves:04}. We constructed a grid covering a large parameter
space, exploring a range of metallicities, ISM densities and
ionization parameters for clusters of ages ranging from 0 to 6 Myr.

\begin{figure*}
\epsscale{1}
\plotone{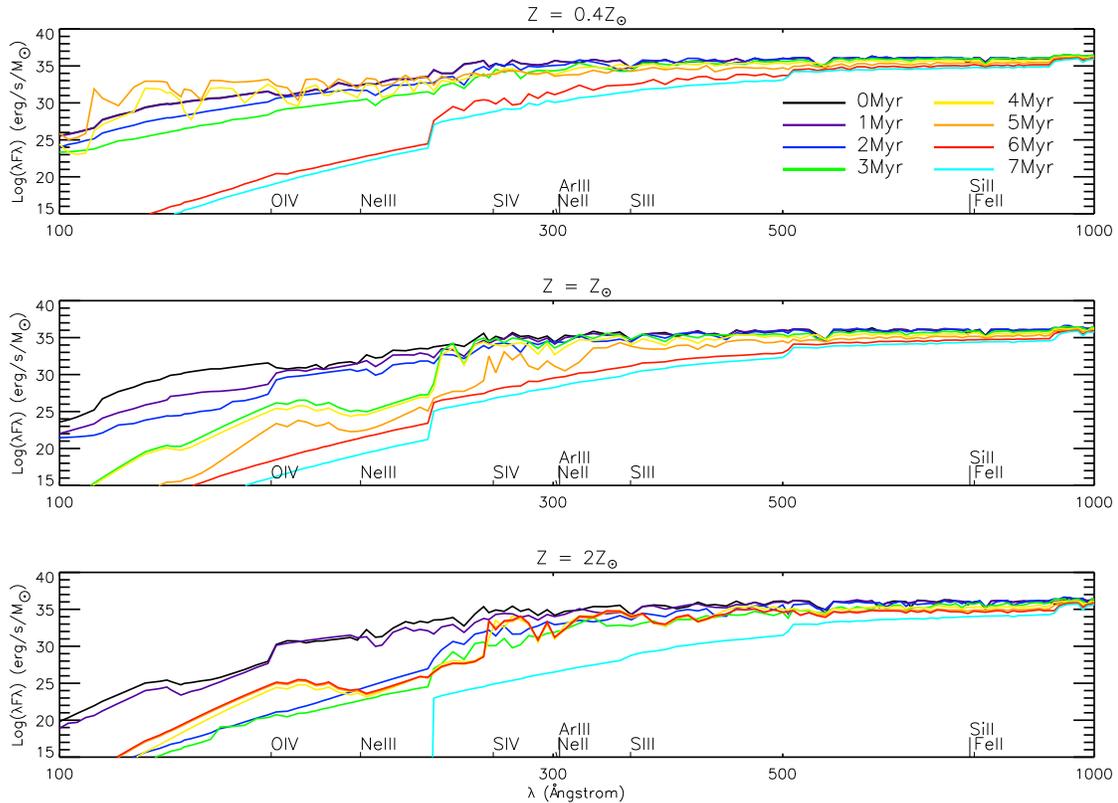}
\caption{FUV spectra as computed by {\it Starburst 99} as a function
of starburst age for {\it upper:} 0.4\Zsol, {\it middle:} 1\Zsol ~and
{\it lower:} 2\Zsol. Ionization thresholds for a number of relevant
elements are shown. 
 \label{UV}}
\end{figure*}

\subsection{Spectral features in the wavelength range of the atmospheric windows}
In the near- and mid-infrared wavelength range a wealth of spectral
features is available to probe the characteristics of the emitting
source, in our case embedded stellar clusters. For these systems, the
near-infrared bands are dominated by bright hydrogen and helium
recombination lines. When the clusters reach the age at which the red
super giant stars appear, these start dominating the near-infrared
continuum and, in the K band, CO absorption bands appear. The spectral
region around 10 \um ~(the N band) displays a complex combination of
silicate absorption, fine-structure emission lines originating from
various species and emission by Polycyclic Aromatic Hydrocarbon (PAH)
molecules. Longwards of the N band, the mid-infrared is dominated by
thermal continuum emission of dust, with fine-structure lines and PAH
emission bands (though much less prominent than in the N band)
superposed on it. Furthermore, in both the near- and mid-infrared
strong molecular hydrogen H$_2$ lines can be found. In this paper we
focus exclusively on the emission lines.

With ground-based observations, information between the atmospheric
windows is unobservable, for instance the \NeIII ~line at 15.56 \um
~between the N and Q band window. For this reason it is impossible to
use \NeIII/\NeII ~ratio, which is a good diagnostic for the hardness
of the radiation field. Longwards of the Q band window there are
several other fine-structure lines unobservable with ground-based
facilities. Most of them are not straightforward to interpret: both
the \OIV ~ and the \FeII ~line around 26 \um ~can be strongly affected
by shocks. The \SiII ~line at 34.82 \um ~is also very complex to
analyse, since it is not only excited in \HII ~regions, but in Photon
Dominated Regions (PDR) and X-ray Dominated Regions (XDR) as well. The
only line with a relatively simple excitation mechanism at these
longer wavelengths accessible to Spitzer is the \SIII ~line at 33.48
\um. Combined with the \SIII ~line at 18.71 \um ~this emission line
forms a very useful density diagnostic, which is unfortunately not
available in ground-based observations.

We will address some of the emission lines only accessible with
space-based telescopes, like the \NeIII ~15.56 \um ~and the \SIII
~33.48 \um ~lines, in order to test our model results with
mid-infrared databases of starburst galaxies (obtained with ISO and
Spitzer, large databases observed with ground-based mid-infrared
facilities are not yet available in the literature). However, the main
focus of this paper is the modeling of ground-based near- and
mid-infrared observations. The lines we have available in these
observations are the series of bright hydrogen and helium
recombination lines in the near-infrared bands plus several
fine-structure lines in the mid-infrared: \ArIII ~at 8.99 \um, \SIV
~at 10.51 \um ~and \NeII ~at 12.81 \um ~in the N band and \SIII ~at
18.71 \um ~in the Q band.

\subsection{Stellar population synthesis model}
We use the recently released version ({\it 5.1}) of {\it Starburst 99}
to model the spectral energy distributions (SEDs) of stellar
clusters. The code constructs these SEDs by summing the
luminosity-weighted spectra of the individual stars according to the
chosen IMF, following theoretical stellar evolutionary tracks for each
star from Zero Age Main Sequence (ZAMS) to its final stage. As we only
study the youngest phases of star formation, the ionizing spectrum is
dominated by O and B stars. The Geneva evolutionary tracks, which are
optimized for modeling high-mass stars, are used to calculate the
SEDs, applying the high mass loss rate models as recommended by
\cite{Maeder:94}. Kurucz's stellar atmosphere models as compiled by
Lejeune \citep{Kurucz:92,Lejeune:97} are applied for stars without
strong winds, for which the static, plane-parallel atmospheres are a
good approximation. For stars with strong winds \cite{Smith:02}
implemented non-LTE, line-blanketed model atmospheres for Wolf-Rayet
\citep[W-R, ][]{Hillier:98} and O stars \citep{Pauldrach:01}. The
switch between extended and plane-parallel atmospheres is as described
in \cite{Leitherer:95}. A detailed description of the changes of
version {\it 5.0} compared to {\it 4.0} can be found in
\cite{Vazquez:05}. Version {\it 5.1} is similar to {\it 5.0}, with
updates in the density and supernova routines\footnote[1]{This
information comes from the {\it Starburst 99} website:
http://www.stsci.edu/science/starburst99/ .}.

We created three sets of models to cover a range in metallicities,
0.4\Zsol, 1\Zsol ~and 2\Zsol (Z=0.008, Z=0.02 and Z=0.04; the solar
abundance set is listed in Table~\ref{tab:elements}). For each of
these metallicities we modeled an instantaneous burst of star
formation, in which a million solar masses of stars are formed
following a Salpeter IMF between 0.1 and 100 \Msol. To ensure that the
chosen cluster mass of 10$^6$ \Msol ~does not induce an artificial
upper mass cutoff, the spectral type distribution of the resulting
stellar populations was examined. At 0 Myr the highest mass bin (O3
star) contains several stars, confirming that the populations fully
sample all spectral types up to 100 \Msol. The spectra are calculated
in 1 Myr time steps between 0 and 6 Myr and additionally at 2.5 and 3.5
Myr. For the 2\Zsol ~models the spectra are calculated at 7 Myr as
well, because of the extended lifetime of the W-R phase at high
metallicity (see Section 3.1).

\begin{figure*}
\epsscale{1}
\plotone{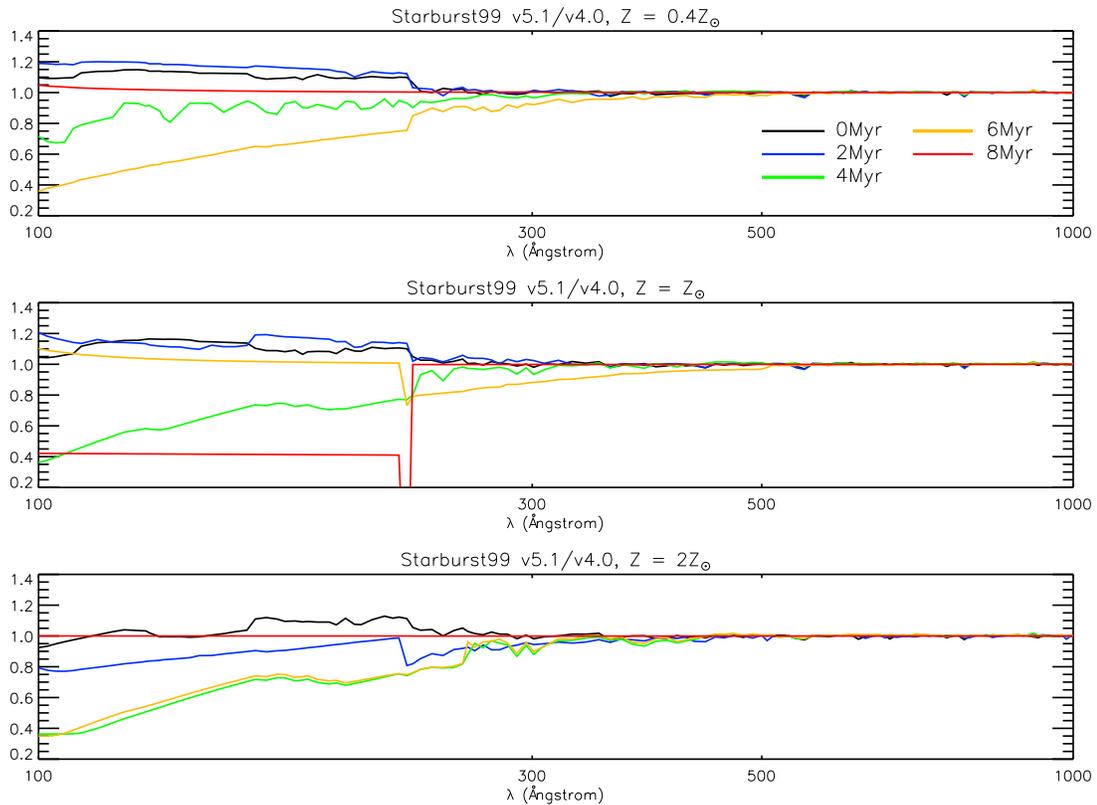}
\caption{Comparison of the FUV spectra computed with the new {\it 5.1
Starburst 99} version and the original {\it 4.0} version. The curves
show the {\it v5.1} spectrum divided by the {\it v4.0} spectrum for
various ages for {\it upper:} 0.4\Zsol, {\it middle:} 1\Zsol ~and {\it
lower:} 2\Zsol. 
 \label{comp_UV}}
\end{figure*}

Figure~\ref{UV} shows the {\it Starburst 99} ionizing spectra
(far-ultraviolet (FUV) between 100 and 1000 \AA) as a function of age
and metallicity. The hardness of the radiation decreases with age,
because the highest mass stars evolve off the main sequence. This
effect is seen most clearly at the shortest wavelengths, between 100
-- 300 \AA. Longwards of 300 \AA ~the SEDs are very similar for all
metallicities, particularly during the first four million years. At
the highest energies (between 100 and 300 \AA) the sub-solar
metallicity spectrum is significantly harder than the super-solar
one. This difference increases with age. Higher metallicity stars
spend a larger fraction of their high energy photons ionizing the
metals in their own atmospheres (line-blanketing). Therefore, fewer
photons are available to ionize the surrounding matter, giving a
softer SED. Furthermore, low metallicity stars of similar spectral
type have a higher effective temperature than high metallicity stars,
also contributing to a harder spectrum at low metallicity. Around 3
Myr the first massive stars enter the W-R phase, causing temporary
hardening of the SED. This manifests itself as a bump around the \OIV
~ionization threshold between 3 and 5 Myr in the \Zsol ~spectra
(between 3 and 6 Myr for 2\Zsol). In the 0.4\Zsol ~spectra the SED
shows a larger number of high energy photons over the whole 100 -- 300
\AA ~range in the W-R phase, making the spectra between 4 and 5 Myr
harder than at 0 Myr. The sharp edge at 228 \AA ~in the spectra at 6/7
Myr corresponds to the \HeII ~ionization edge (54.42 eV). At these
ages there are few photons available for the excitation of species
with an excitation potential exceeding that of \HeII ~(like \NeIII
~and \OIV).

The FUV spectra have changed significantly since the {\it Starburst 99
  4.0} version. Figure~\ref{comp_UV} shows the effect of the
implementation of sophisticated atmospheric models, including metal
line-blanketing and non-LTE O-star atmospheres ({\it v4.0} only
treated radiative transfer for hydrogen and helium and assumed LTE for
stars of all masses). The plots show the new {\it Starburst 99 v5.1}
spectra divided by the output of the old version of the code ({\it
  v5.1/v4.0}). The largest differences occur shortwards of 300 \AA. At
the youngest ages (0 -- 2 Myr for 0.4\Zsol ~and \Zsol ~and at 0 Myr
for 2\Zsol) the new version produces 5 -- 20\% more \HeII ~ionizing
photons ($<$ 300 \AA) than {\it v4.0}. Between 2 -- 6 Myr, the FUV
spectra ($<$ 300 \AA) of the old version harden where the output
spectra of the new version show a gradual softening with age, as
discussed above. The relative amount of high energy photons available
in the {\it v5.1} spectra between 2 -- 6 Myr can be as much as 20\% (at
250 \AA) to 60\% (at 100 \AA) lower compared to the old version. This
shows that a considerable fraction of photons at these energies is
used to ionize the stellar atmospheres. Around 6 or 7 Myr (depending
on metallicity) the old and new version give roughly identical output
spectra. To conclude, characteristics of young stellar populations
and/or their surroundings derived from emission lines sensitive to
emission in this wavelength range alter significantly with these
recent improvements of the code.

\subsection{Photoionization model}

\begin{deluxetable}{lll}
  \tabletypesize{\footnotesize} \tablecaption{Solar abundance \& metallicity scaling \label{tab:elements}}
  \tablewidth{0pt} 
\tablehead{ \colhead{Element} &
            \colhead{Abundance\tablenotemark{a}} &
            \colhead{Depletion\tablenotemark{b}} 
}

\startdata

&&\\
H .................................. & \phs0.000\phn   & \phs0.00\phn\\
He.................................. & $-$1.01\phn    & \phs0.00\phn\\
C .................................. & $-$3.59\phn\phn & $-$0.30\phn \\
N .................................. & $-$4.20\phn\phn & $-$0.22\phn \\
O .................................. & $-$3.34\phn\phn & $-$0.07\phn \\
Ne.................................. & $-$3.91\phn\phn & \phs0.00\phn\\
Na.................................. & $-$5.75\phn\phn & $-$1.00\phn \\
Mg.................................  & $-$4.42\phn\phn & $-$0.70\phn \\
Al.................................. & $-$5.61\phn\phn & $-$1.70\phn \\
Si...................................& $-$4.49\phn\phn & $-$1.00\phn \\
S .................................. & $-$4.79\phn\phn & \phs0.00\phn\\
Cl.................................. & $-$6.40\phn\phn & $-$1.00\phn \\
Ar.................................  & $-$5.20\phn\phn & \phs0.00\phn\\
Ca.................................  & $-$5.64\phn\phn & $-$2.52\phn \\
Fe.................................. & $-$4.55\phn\phn & $-$2.00\phn \\
Ni.................................. & $-$5.68\phn\phn & $-$2.00\phn \\

  \enddata
\tablenotetext{a}{All abundances are logarithmic with respect to Hydrogen}
\tablenotetext{b}{Depletion given as
$\log(X/\mathrm{H})_{\mathrm{gas}}-
\log(X/\mathrm{H})_{\mathrm{ISM}}$}

\end{deluxetable}

\begin{figure*}
\epsscale{1}
\plotone{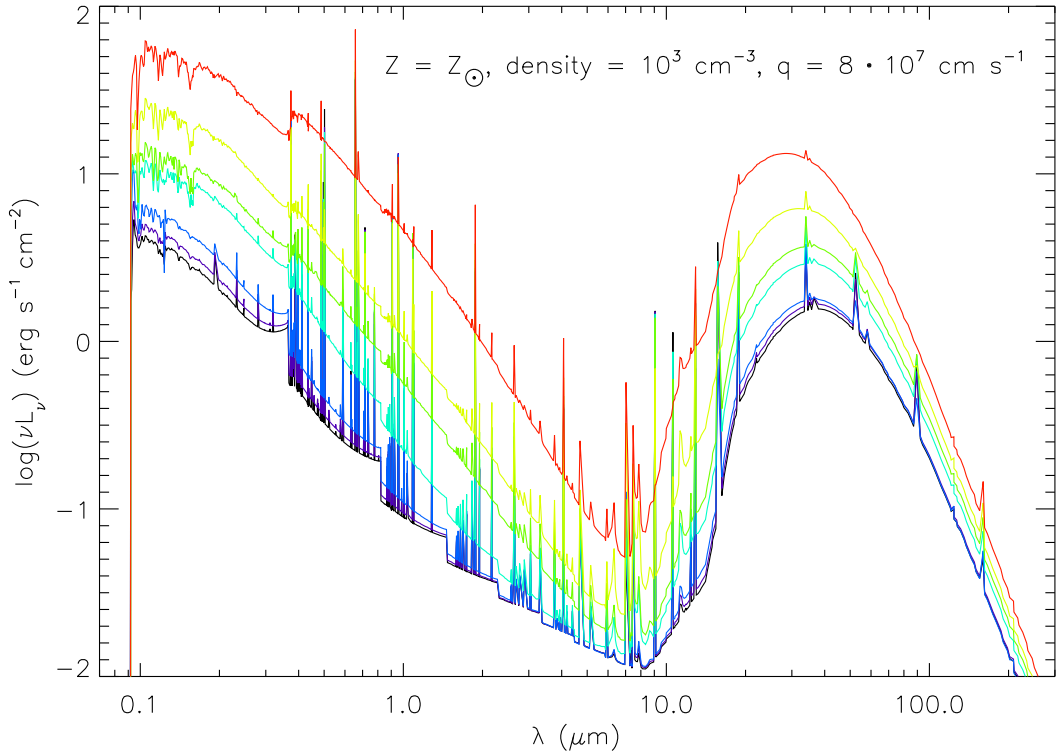}
\caption{Example of {\it Mappings} output spectra. Solar metallicity
model with a density of 10$^3$ cm$^{-3}$ and an ionization parameter
$q$ of 8 $\cdot$ 10$^7$ cm s$^{-1}$. The age increases from {\it
lowest:} 0 Myr to {\it upper:} 6 Myr, in steps of 1 Myr (color version
available in electronic edition). Note the nebular bound-free edges,
the strong recombination and fine-structure emission lines and the
far-infrared peak originating from heated dust.
 \label{output_spec}}
\end{figure*}

The {\it Starburst 99} output spectra are fed into the photoionization
code {\it Mappings} to model the combined SED of a stellar cluster
plus the irradiated surrounding gaseous medium. The code combines
photoionization theory with radiation pressure and the effects of
dust. The cloud's ionization structure is determined by the ionization
parameter $q$ at the inner cloud boundary:

\begin{equation}
\label{eq:q}
q = \frac{Q_{\rm Lyc}}{4\pi R^2n_{\rm ion}} 
\end{equation}

$Q_{\rm Lyc}$ being the hydrogen ionizing photon flux, $R$ the radius
of the inner cloud boundary, and $n_{\rm ion}$ the ion density (which
equals $\approx$ 1.1 times the hydrogen density $n_H$ if we account
for helium ionization). The ionization parameter $q$ measures the
density of ionizing photons relative to the atomic density. It relates
to the commonly used dimensionless ionization parameter U through U
$\equiv$ 1.1 $\cdot$ $q/c$. We assume that the object's radius $R$ is
much larger than the thickness of the emitting shell of gas. In this
case a plane-parallel geometry is a good approximation, which we
applied in our models. As the code steps through the interstellar
cloud, {\it Mappings} calculates the energy balance for each thin slab
by coupling detailed modeling of the atomic gas processes with the
dust physics.  Photoionization, collisional ionization and
recombination processes are taken into account, as well as
photoelectric heating of the gas by the dust, radiation pressure
acting on the dust and charge transfer reactions. In this way the
(FUV) photons from the input spectrum are reprocessed by the ISM
surrounding the star cluster. For each given combination of input SED,
metallicity, density and ionization parameter the photoionization code
gives a unique emission line spectrum (see Fig.~\ref{output_spec} and
a detailed description in Section 3).

The dust is modeled by a combination of two types of grains, amorphous
silicates and amorphous carbon grains. Together these adequately
reproduce both continuum emission in the infrared and extinction at
shorter wavelengths \citep{Draine:03, Dopita:05}. The grain sizes
follow a power law distribution with a slope consistent with the grain
shattering model \citep{Jones:96}, $\alpha$ $=$ -3.3, modified to
allow for a lower and upper size cutoff of 40 \AA ~and 1600 \AA
~respectively \citep[see Equation 19 in][]{Dopita:05}. {\it Mappings}
divides the grains into 80 size bins and computes the absorption,
scattering, and photoelectric heating for each bin (in older versions
of the code photoelectric heating was calculated for the dust as a
whole). The results are then used to calculate the dust temperature,
which determines the mid- and far-infrared continuum emission. While
only two simple types of dust are considered in the calculation of
dust extinction and emission, all elements with a high enough gas
phase condensation temperature ($>$ few 100 K) are depleted. Depletion
factors are assumed to be identical to those in the local ISM
\citep{Savage:96}. The depletion set is listed in
Table~\ref{tab:elements}.

Although PAH features have been observed routinely ever since their
discovery two decades ago, exact knowledge on PAH shapes, size
distributions and their interplay with the environment still remains
too poor for PAHs to be modeled adequately. In {\it Mappings} PAHs are
represented by a single type, coronene ($\rm C_{24}H_{12}$), for which
the abundance is determined by the amount of carbon depleted onto PAH
molecules (fraction here chosen to be 0.05). For this one specific PAH
molecule absorption, emission, and photoelectric processes are
calculated in detail to represent the PAH infrared spectrum
\citep[][and references therein]{Dopita:05}.

\subsection{Parameter space covered by the model grid}

The model input parameters for {\it Mappings} were chosen to mimic the
wide spectrum of observed nebular conditions in \HII ~regions,
covering low-density to UC\HII ~regions. For a range in ages (0 -- 6/7
Myr) and metallicities (0.4\Zsol, \Zsol ~and 2\Zsol) we computed
plane-parallel, isobaric, dusty models with gas densities increasing
from $10^2$ to $10^6$ cm$^{-3}$. Likewise an appropriate range in
values for the the ionization parameter $q$ (see Eq. \ref{eq:q}) was
chosen. Typical values for the ionization parameter in local starburst
galaxies span $-$3 $\le$ logU $\le$ $-$1.5 \citep[Fig.~10 in][who
adopt a value for logU of $-$2.3 in their models]{Rigby:04}. Therefore
the {\it Starburst 99} spectra are scaled to obtain a fixed ionization
parameter $q$ at the inner cloud boundary at a value of 2 $\cdot$
10$^7$, 4 $\cdot$ 10$^7$, 8 $\cdot$ 10$^7$, 1.6 $\cdot$ 10$^8$, 4
$\cdot$ 10$^8$ and 8 $\cdot$ 10$^8$ cm s$^{-1}$(corresponding to logU
$\approx$ $-$3.13, $-$2.83, $-$2.53, $-$2.23, $-$1.83 and
$-$1.53). Scaling the {\it Starburst 99} spectra essentially means
scaling the cluster mass. We will not address the cluster masses in
present work, but we will use the mass information in future work on
near-IR spectra of the same objects (Snijders \& Van der Werf, in
preparation).

\section{Model results}

Fig.~\ref{output_spec} shows an example of the {\it Mappings} output
spectra. The UV part of the spectrum originates from massive stars at
all times. The visible and near-infrared are dominated by nebular
emission up to 4 Myr, showing clear bound-free edges. Beyond 4 Myr red
super giants take over the near-infrared continuum. Additionally,
prominent hydrogen and helium recombination lines are found in the
visible and near-infrared. The mid- and far-infrared continuum
originates from heated dust, showing the characteristic far-infrared
peak. In the mid-infrared N band around 10 \um ~the SED is dominated
by a combination of fine-structure lines, silicate absorption and PAH
features.

The peak of the hot dust continuum is seen to shift towards shorter
wavelengths, indicating higher dust temperatures, with increasing age
(and thus softer ionizing spectrum). This is caused by the way we
define the {\it Mappings} input parameters. The {\it Starburst 99}
input spectra, pressure and temperature are scaled to keep the density
and ionization parameter at certain fixed values. A softer input
spectrum requires a larger scaling factor to obtain the same amount of
ionizing photons. This results automatically in a larger number of
photons in the soft-UV and visible regime with increasing age. These
photons do not affect the mid-infrared fine-structure lines, since
these lines are exclusively excited by high energy UV photons. The
dust however is heated by both. Altogether, for a fixed value of the
ionization parameter, there are more photons available at higher age
to heat the dust, causing the far-infrared peak to shift to shorter
wavelengths.

\subsection{Age evolution of line ratios: mid-infrared}

\begin{figure*}
\epsscale{1}
\plotone{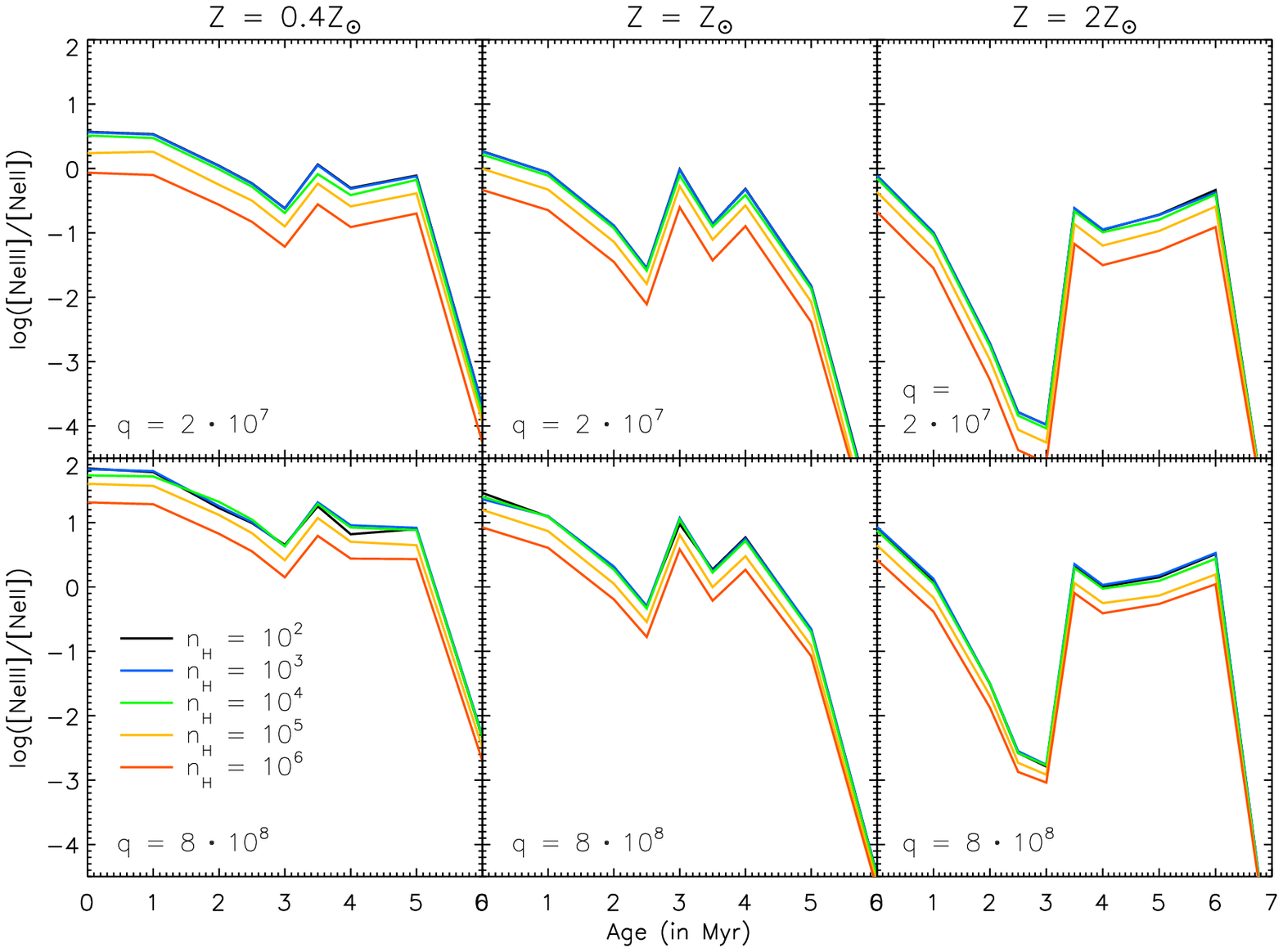}
\plotone{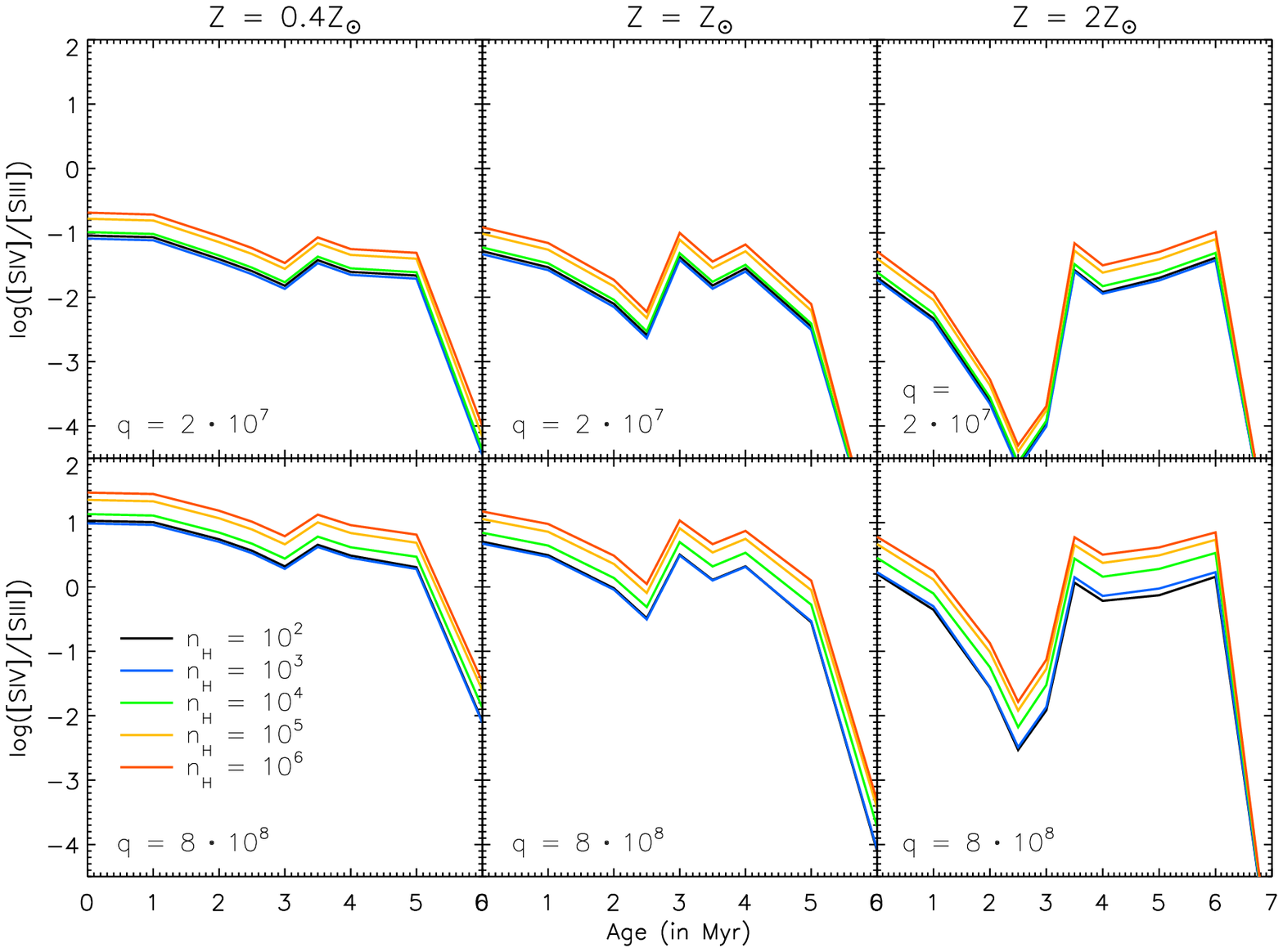}
\caption{Evolution of the emission line ratios with cluster age of
{\it upper}: \NeIII$_{\rm 15.56 \mu m}$/\NeII$_{\rm 12.81 \mu m}$ and
{\it lower}: \SIV$_{\rm 10.51 \mu m}$/\SIII$_{\rm 18.71 \mu m}$. The
left panels show the curves for Z = 0.4\Zsol, the middle panels Z = \Zsol ~and
the right panels Z = 2\Zsol. The upper row shows the case of low
ionization parameter, $q$ = 2 $\cdot$ 10$^7$ cm s$^{-1}$, the lower
row that of high ionization parameter, $q$ = 8 $\cdot$ 10$^8$ cm
s$^{-1}$. Curves of different line style (different color in electronic
edition) represent results of models with different densities, ranging
from 10$^2$ cm$^{-3}$ to 10$^6$ cm$^{-3}$. Line ratio values can be
found tabulated at www.ifa.hawaii.edu/$\tilde{\ }$kewley/Mappings/IRdiagnostics.
\label{AgeEvol1}}
\end{figure*}

\begin{figure*}
\epsscale{1}
\plotone{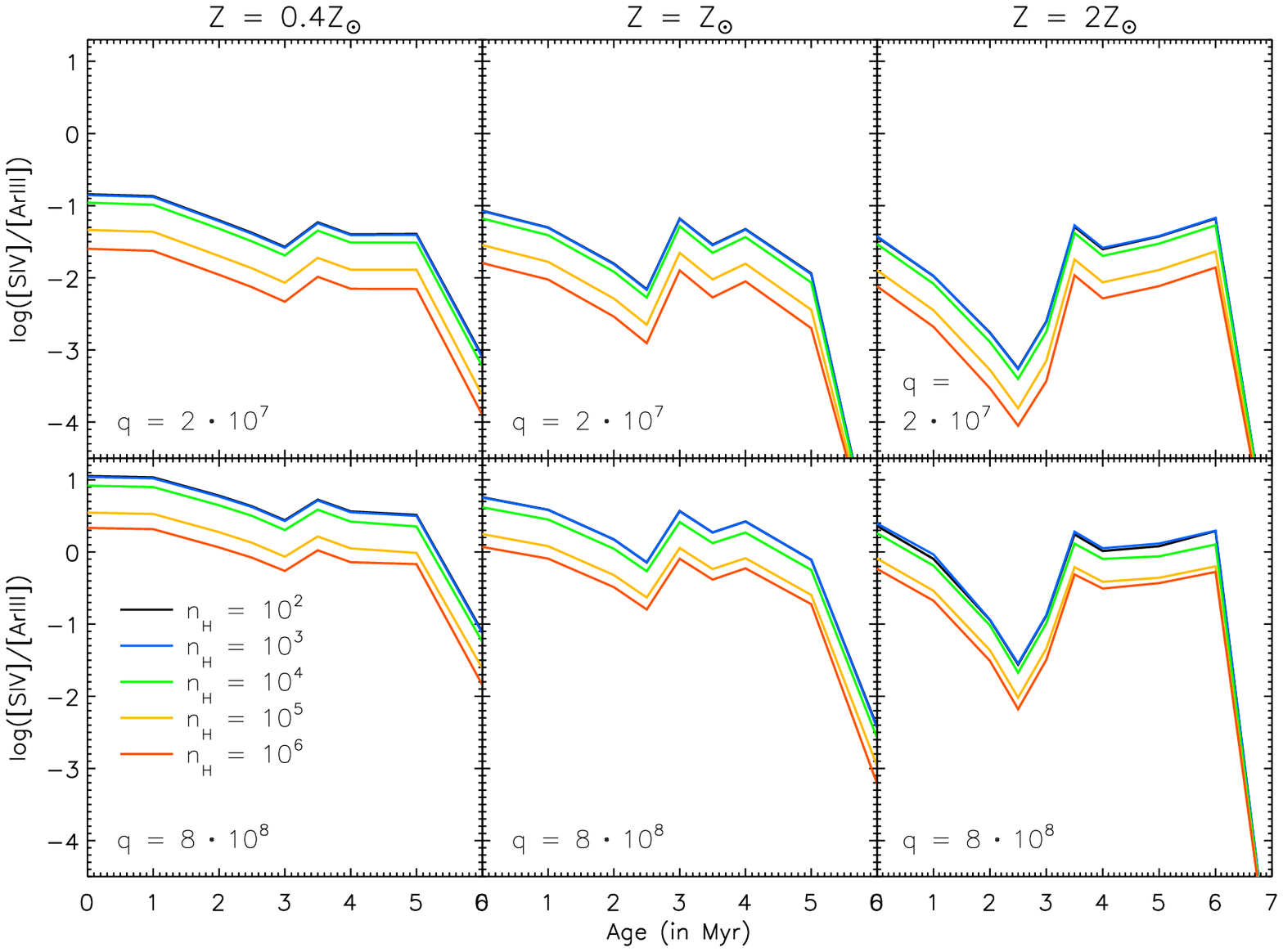}
\plotone{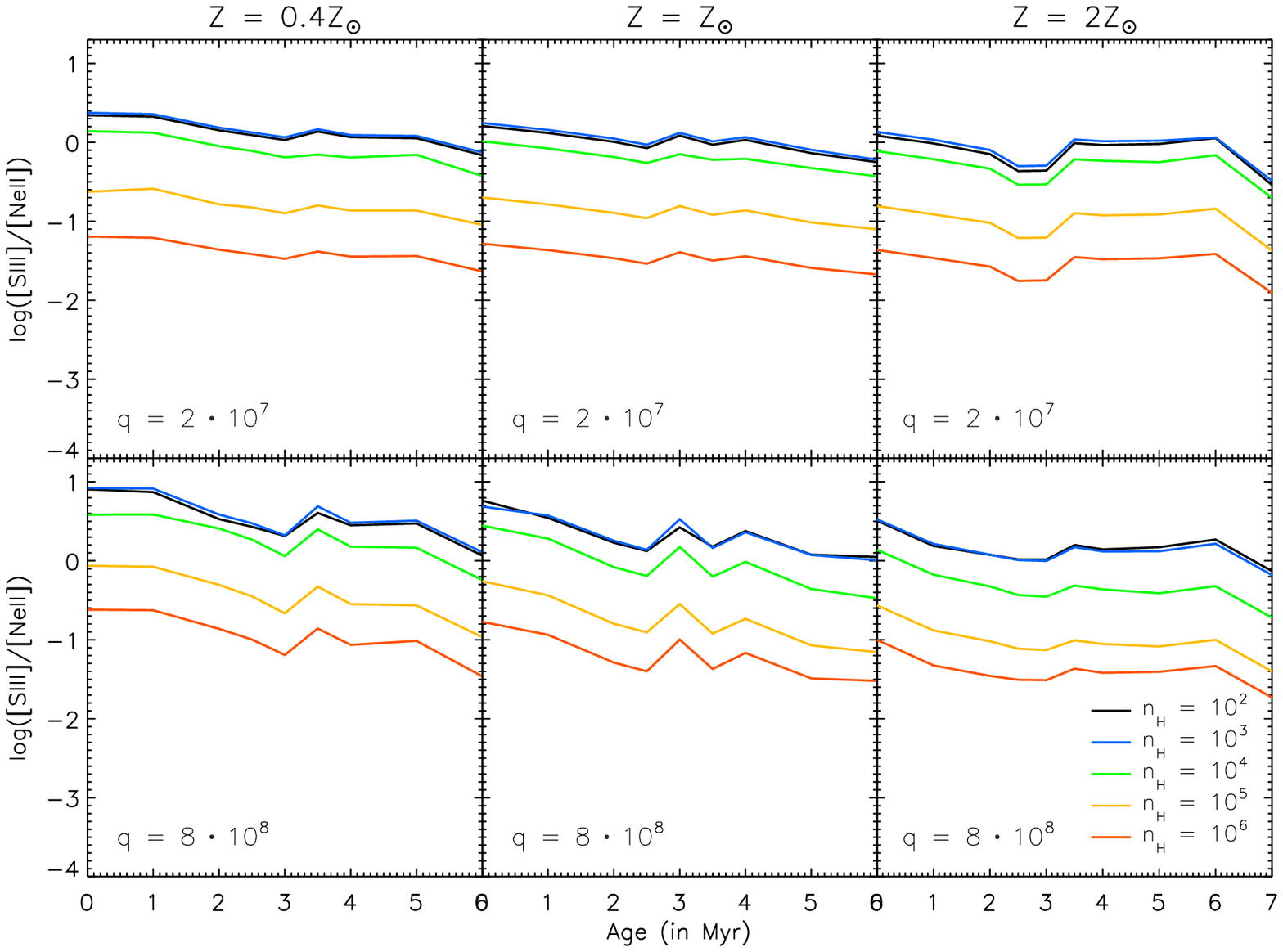}
\caption{Like Fig.~\ref{AgeEvol1} for {\it upper}: \SIV$_{\rm 10.51
\mu m}$/\ArIII$_{\rm 8.99 \mu m}$ and {\it lower}: \SIII$_{\rm 18.71
\mu m}$/\NeII$_{\rm 12.81 \mu m}$.
\label{AgeEvol2}}
\end{figure*}

Ratios of lines with different excitation potentials can be used to
measure the temperature of the radiation field. Unfortunately the
interpretation of these line ratios is complicated by the fact that
they are sensitive to metallicity of the ionizing cluster, ISM density
and ionization parameter as well. This effect is demonstrated in
Figures \ref{AgeEvol1} and \ref{AgeEvol2}, in which
the evolution of various emission line ratios with cluster age is
shown.  Line ratios involving emission lines that have (almost)
identical excitation potentials, but different critical densities, are
useful as density probes. 

To be able to compare our model results with models from the
literature we discuss the temperature-sensitive ratios \NeIII$_{\rm
15.56 \mu m}$/\NeII$_{\rm 12.81 \mu m}$ ~and \SIV$_{\rm 10.51 \mu
m}$/\SIII$_{\rm 18.71 \mu m}$ and the density-sensitive ratio
\SIII$_{\rm 33.48 \mu m}$/ \SIII$_{\rm 18.71 \mu m}$, even though some
of these ratios involve lines observable from space only. If we focus
on the emission lines that are observable from the ground, we are
limited to \ArIII$_{\rm 8.99 \mu m}$, \SIV$_{\rm 10.51 \mu m}$ ~and
\NeII$_{\rm 12.81 \mu m}$ ~in the N band and \SIII$_{\rm 18.71 \mu m}$
~in the Q band. So, in addition we examine the 'ground-based' line
ratios temperature-sensitive ratio \SIV$_{\rm 10.51 \mu
m}$/\ArIII$_{\rm 8.99 \mu m}$ ~and the density-sensitive ratio
\SIII$_{\rm 18.71 \mu m}$/\NeII$_{\rm 12.81 \mu m}$.

The line ratios are later combined to form diagnostic ratio--ratio
plots (see Section 5). First, \SIV/\SIII ~ is plotted versus
\NeIII/\NeII ~to construct the commonly used diagnostic diagram
sensitive to the hardness of the ionizing radiation. Secondly, the
emission lines that are accessible from the ground are combined to
form an extinction independent diagnostic diagram, plotting \SIV/\ArIII
~versus \SIII/\NeII. Adopting a standard \cite{Draine:89} extinction
curve, the intrinsic values of \SIV/\ArIII ~and \SIII/\NeII ~differ
from the observed ratios by less than 3$\%$, which is much less than
the calibration uncertainties of our observations. These ratios are
thus essentially unaffected by reddening.

Note that all mid-infrared fine-structure lines under study here are
generated by $\alpha$--elements (neon, argon and sulphur), so
abundance ratios are expected to be approximately fixed. All
differences in the line ratio behavior with metallicity originate
from other processes. \linebreak

As the stellar population ages and massive stars evolve off the main
sequence, the FUV spectrum softens causing ratios of high to low
ionization lines like \NeIII/\NeII ~to drop. The most massive star
present at 2.5 Myr is an O5.5 star. At later ages the line ratios show
an upturn corresponding to the appearance of W-R stars, which dominate
the radiation field until the age of 6 Myr (see Figs.~\ref{AgeEvol1} and
\ref{AgeEvol2}).

High metallicity stars have softer FUV spectra than low metallicity
stars of similar mass due to line-blanketing within the stellar
atmospheres. This explains the offset towards lower values in the
\NeIII/\NeII ~ratio with increasing metallicity as well as the steeper
drop in the ratio in the first 2 -- 3 Myr (see upper panel in
Fig.~\ref{AgeEvol1}). After exhaustion of hydrogen burning in the
core, stellar atmosphere temperatures can drop to low enough values to
initiate the W-R phase of high mass loss. Apart from having softer
spectra initially, metal-rich stars can cool more efficiently, causing
their W-R phase to start earlier than that of metal-poor stars. The
first W-R stars appear at 2 Myr for 2\Zsol , 2.5 Myr for \Zsol and 3
Myr for 0.4\Zsol. Furthermore, the phase in which W-R stars dominate
the FUV spectrum lasts longer in the high metallicity case. Since
metals are more abundant in their stellar atmospheres, lower mass
metal-rich stars can generate enough radiation pressure to enter the
W-R phase, where metal-poor stars of similar mass cannot. This effect
extends the W-R phase to higher ages in the 2\Zsol ~population, which
is apparent from high values for \NeIII/\NeII ~up to 6 Myr.

Given the values of the critical densities \citep[1.8 $\cdot$ 10$^5$
and 6.1 $\cdot$ 10$^5$ cm$^{-3}$ for \NeIII$_{\rm 15.56 \mu m}$ ~and
\NeII$_{\rm 12.81 \mu m}$ ~respectively; all values for critical
densities are obtained from][]{Tielens:05} the \NeIII/\NeII ~is mostly
sensitive to densities between 10$^4$ and 10$^6$ cm$^{-3}$. The ratio
decreases with approximately a factor of three if density increases
from 10$^4$ to 10$^6$ cm$^{-3}$. Furthermore, the \NeIII/\NeII ~ratio
is sensitive to the ionization parameter, because a larger value of
$q$ means that more ionizing photons are available per atom. The
\NeIII/\NeII ~ratio increases more than an order of magnitude when $q$
increases from 2 $\cdot$ 10$^7$ to 8 $\cdot$ 10$^8$ cm
s$^{-1}$. \linebreak

\begin{figure*}
\epsscale{1}
\plotone{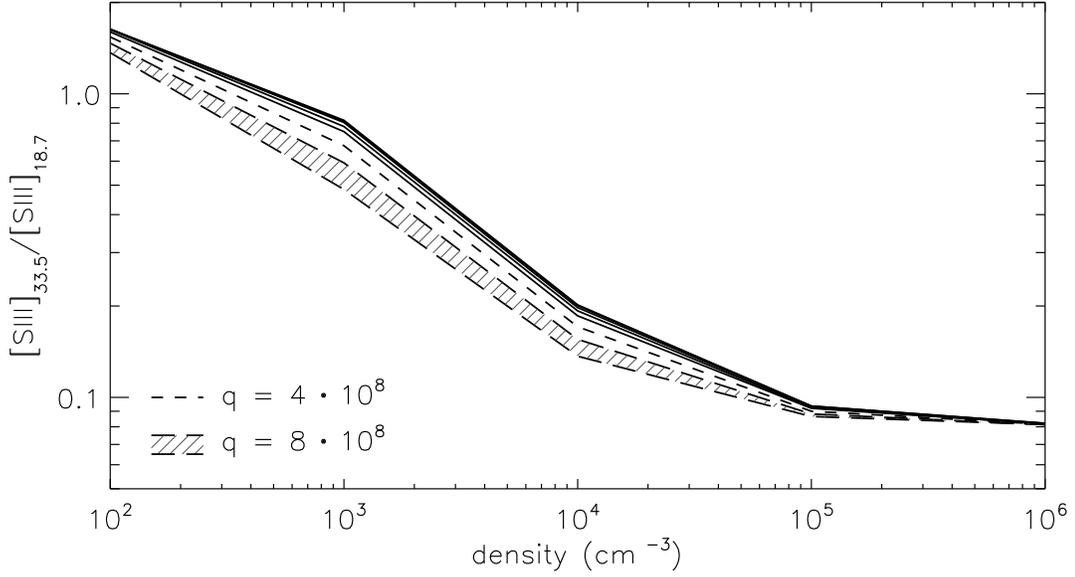}
\caption{Evolution of the \SIII$_{\rm 33.48 \mu m}$/\SIII$_{\rm 18.71
\mu m}$ ~ratio with increasing density. The solid lines result from
the models with an ionization parameter of 2 $\cdot$ 10$^7$, 4 $\cdot$
10$^7$, 8 $\cdot$ 10$^7$ and 1.6 $\cdot$ 10$^8$ cm s$^{-1}$ (top to
bottom solid line, the 2 $\cdot$ 10$^7$ and 4 $\cdot$ 10$^7$ are
indistinguishable). For a higher ionization parameter the curve drops
towards somewhat lower \SIII/\SIII ~values (short-dashed line: 4
$\cdot$ 10$^8$ cm s$^{-1}$; long-dashed line: 8 $\cdot$ 10$^8$ cm
s$^{-1}$ ; the shaded area connects the $q$ = 8 $\cdot$ 10$^8$ cm
s$^{-1}$ curves of different ages, the upper curve represents the
average values for 0 -- 4 Myr and the lower curve shows the 5 Myr
values). The differences between the curves is caused by a weak
dependence of the \SIII/\SIII ~ratio on electron temperature ~(see
Section 3.1 for explanation). Line ratio values can be found tabulated
at www.ifa.hawaii.edu/$\tilde{\ }$kewley/Mappings/IRdiagnostics.
\label{SIIISIII}}
\end{figure*}

\begin{figure*}
\epsscale{1}
\plotone{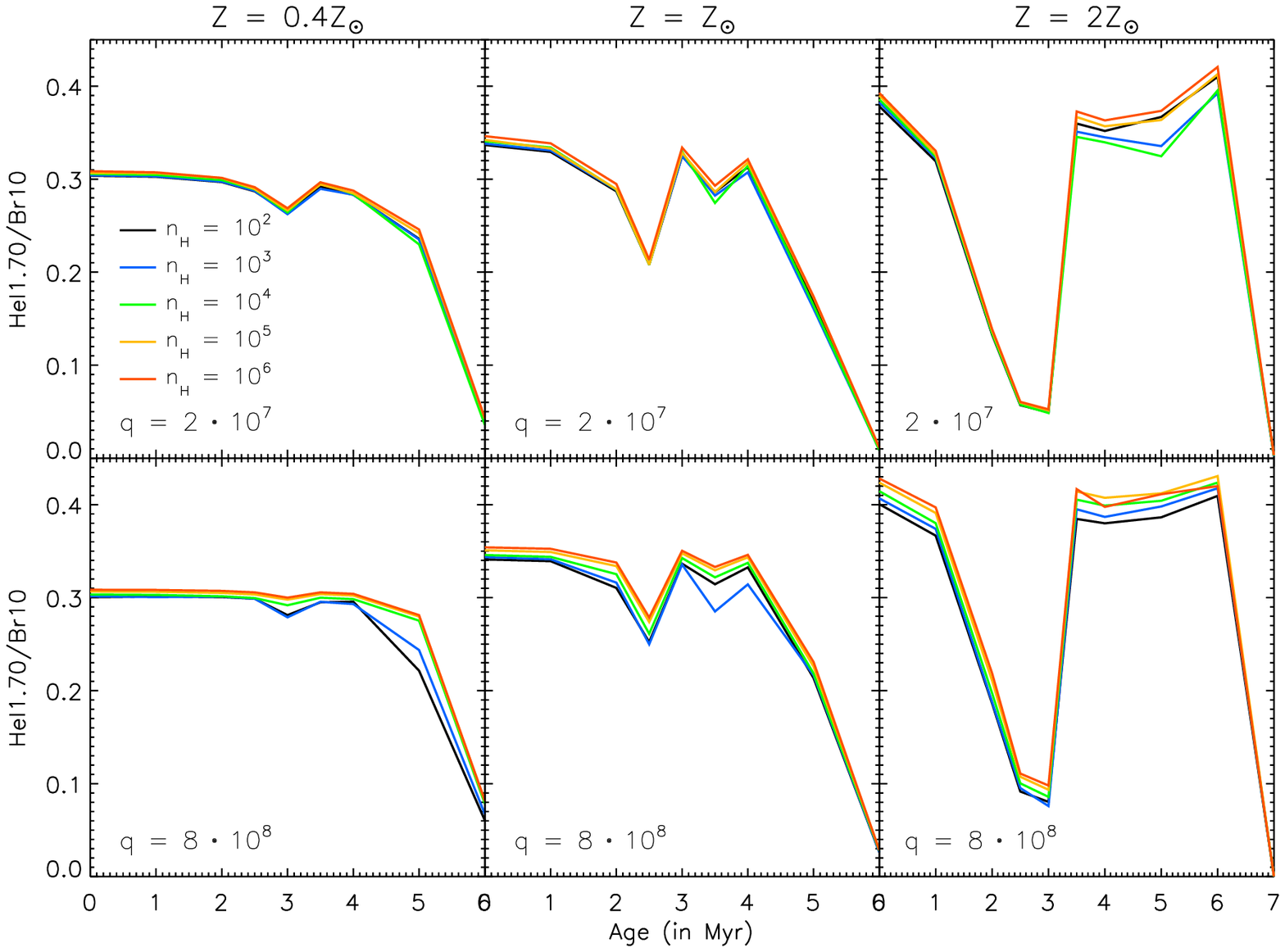}
\caption{Like Fig.~\ref{AgeEvol1} for  \HeI$_{\rm 1.70 \mu m}$/Br10.
\label{AgeEvol3}}
\end{figure*}

\begin{figure*}
\epsscale{1}
\plotone{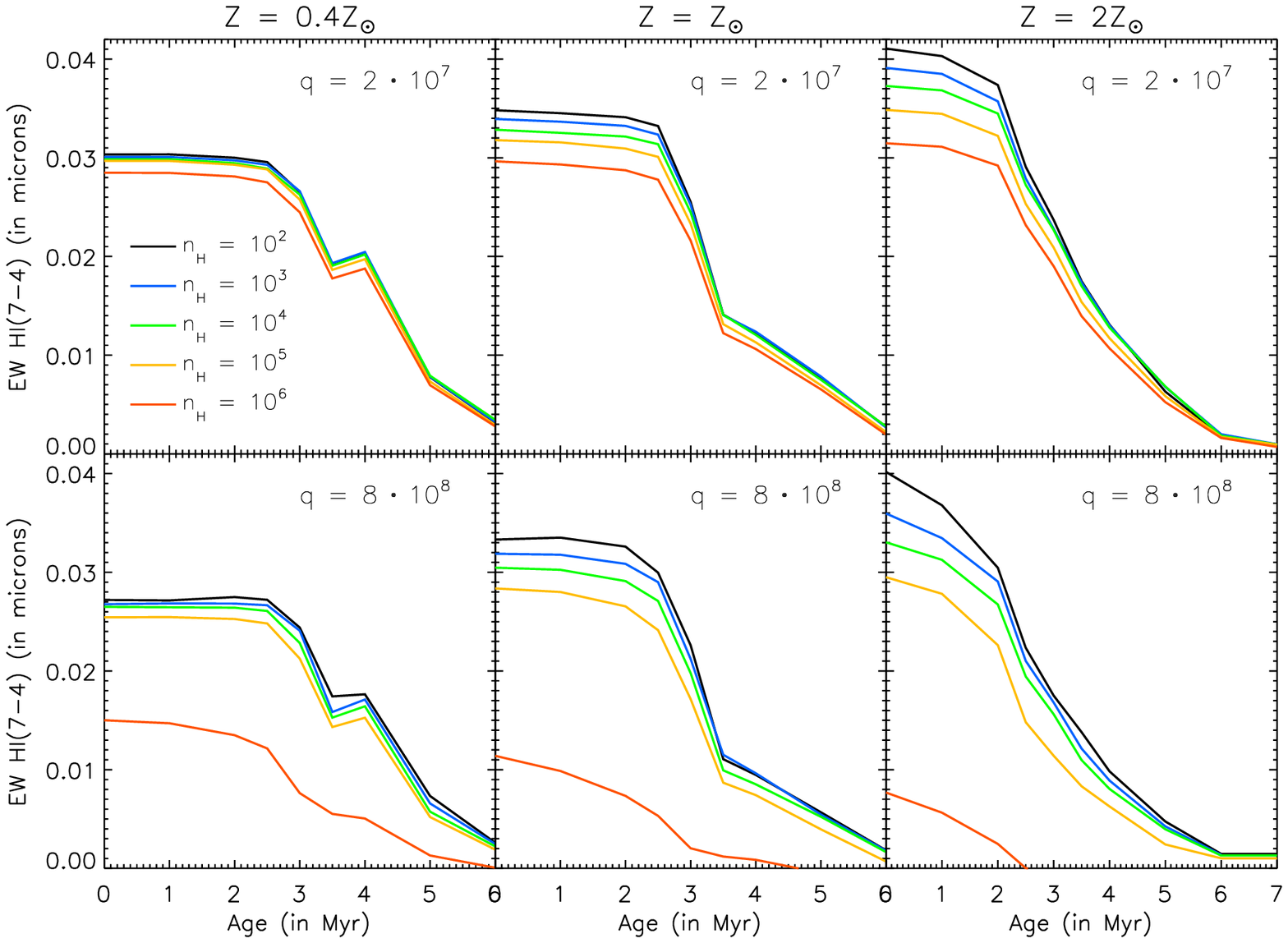}
\caption{Age evolution of the equivalent width of the hydrogen
 recombination line Br$\gamma$. The left panels show the curves for
 Z = 0.4\Zsol, the middle panels Z = \Zsol and the right panels Z = 2\Zsol. The
 upper row shows the case for low ionization parameter, $q$ = 2
 $\cdot$ 10$^7$ cm s$^{-1}$, the lower row that of high ionization
 parameter, $q$ = 8 $\cdot$ 10$^8$. Curves of different line style
 (different color in electronic edition) represent results of models
 with different density, ranging from 10$^2$ cm$^{-3}$ to 10$^6$
 cm$^{-3}$. Line ratio values can be found tabulated at
 www.ifa.hawaii.edu/$\tilde{\ }$kewley/Mappings/IRdiagnostics.
\label{EW}}
\end{figure*}

The evolution of \SIV/\SIII ~shows similar behavior as \NeIII/\NeII,
with subtle differences due to (1) the specific sensitivity of the
various individual lines to the spectral hardness of the radiation and
(2) the density of the radiating medium (lower panel in
Fig.~\ref{AgeEvol1}). With critical densities of 3.7 $\cdot$ 10$^4$
cm$^{-3}$ for \SIV ~and 1.0 $\cdot$ 10$^4$ cm$^{-3}$ for \SIII,
\SIV/\SIII ~is sensitive to slightly lower densities than
\NeIII/\NeII. The density dependence is reversed, showing an
increasing line ratio with increasing density. The sensitivity to $q$
is stronger for \SIV/\SIII ~than for \NeIII/\NeII. As $q$ changes from
2 $\cdot$ 10$^7$ to 8 $\cdot$ 10$^8$ cm s$^{-1}$ the \SIV/\SIII ~line
ratio increases almost two orders of magnitude . \linebreak

Inspecting the evolution of \SIV/\ArIII, one again recognizes the same
shape as for \NeIII/\NeII ~and \SIV/\SIII ~(see
upper panel Fig.~\ref{AgeEvol2}). \SIV/\ArIII ~is equally sensitive to $q$ as
\SIV/\SIII ~and somewhat more sensitive to density, because the
difference in critical densities between \SIV ~and \ArIII ~is larger
than between \SIV ~and \SIII ~(3.7 $\cdot$ 10$^4$ and 3.7 $\cdot$
10$^5$ cm$^{-3}$ for \SIV ~and \ArIII ~respectively). \linebreak

Proceeding to the density-sensitive line ratios, we note that
\SIII$_{\rm 33.48 \mu m}$/\SIII$_{\rm 18.71 \mu m}$ ~is most commonly
used in the mid-infrared as a density estimator. Having the same
excitation potential, the ratio of the double-ionized sulphur lines is
mainly affected by the difference in collisional de-excitation
rate. With critical densities of 1.2 $\cdot$ 10$^3$ and 1.0 $\cdot$
10$^4$ cm$^{-3}$ for the \SIII ~33.48 \um ~and the 18.71 \um ~line
respectively, the \SIII/\SIII ~ratio is most sensitive in the density
range between 10$^2$ and 10$^4$ cm$^{-3}$, as can be seen in
Fig.~\ref{SIIISIII}. Note that the sulphur line ratio changes lightly
with $q$. This effect is caused by the fact that critical densities
are mildly sensitive to electron temperature. So, in general,
density-sensitive line ratios are weakly dependent on electron
temperature as well; lines with a higher critical density favor a
higher temperature. Nebulae characterized by high ionization
parameters have high average electron temperatures. This effect causes
the decrease of \SIII/\SIII ~towards lower values with increasing
ionization parameter.

The excitation potentials of \SIII ~and \NeII ~(23.34 eV and 21.56 eV
respectively) are close enough to make the ratio almost insensitive to
the hardness of the radiation field. The substantially different
critical densities, 6.1 $\cdot$ 10$^5$ cm$^{-3}$ for \NeII ~and 1.0
$\cdot$ 10$^4$ cm$^{-3}$ for \SIII, make \SIII/\NeII ~very sensitive
to the ISM density in the 10$^4$ -- 10$^6$ cm$^{-3}$ range (lower
panel in Fig.~\ref{AgeEvol2}). Being sensitive to a different density
regime \SIII/\NeII ~is not so much a replacement for \SIII/\SIII ~as a
density probe, but nicely complementary.

\subsection{Age evolution of line ratios: near-infrared}

If we expand towards shorter wavelengths and include the near-infrared
regime in our analysis, several other useful diagnostics become
available. For one, the near-infrared is rich in hydrogen and helium
recombination lines. As with the mid-infrared fine-structure lines
these near-infrared lines can be used as a thermometer. In a soft
radiation field, there are many more photons available for the
excitation of hydrogen, which has the lowest potential (13.6 eV for
hydrogen versus 24.6 eV for helium), resulting in weak \HeI
~recombination lines relative to the \HI ~lines. With an increasingly
hard radiation field the helium Str\"omgren sphere gradually fills
that of hydrogen, observable by a higher ratio, until it finally
saturates when the two Str\"omgren spheres coincide \citep[Fig.~5
in][]{Thornley:00}.

\HeI$_{\rm 2.06 \mu m}$/Br$\gamma$ ~has been shown to be an unreliable
diagnostic, or at least one that is very complicated to interpret
\citep{Rigby:04, Lumsden:01}. This is because the \HeI$_{\rm 2.06 \mu
m}$ ~2$^1$P -- 2$^1$S line does not purely arise from a recombination
cascade. The population of the 2$^1$P level can be affected both by
collisional pumping from the metastable 2$^1$S level as well as
resonance scattering of the \HeI ~2$^1$P -- 2$^1$S transition at 584
\AA. This effect makes the helium line very sensitive to nebular
conditions. The \HeI$_{\rm 1.70 \mu m}$ ~originates almost exclusively
from recombination cascade, making a \HeI/\HI ~ratio involving this
line a much cleaner diagnostic. \HeI$_{\rm 1.70 \mu m}$ and \HI(10--4)
(Brackett 10, Br10) are close in wavelength and \HeI$_{\rm 1.70 \mu
m}$/Br10 is therefore hardly affected by reddening. This ratio has
been extensively tested as stellar temperature diagnostic and found to
show good agreement with other temperature indicators
\citep{Rigby:04}. Unfortunately, the \HeI$_{\rm 1.70 \mu m}$ line is
relatively faint, less than one-tenth of the Br$\gamma$
intensity. This means it can usually only be detected in nearby
objects with very bright recombination lines.

{\it Mappings} does not predict the intensity of either the \HeI ~2.06
or the 1.70 \um ~line. To obtain the \HeI ~1.70 \um ~line strength we
therefore scale the \HeI ~4471 \AA ~line, which originates from the
same upper level. Under the assumption of case B recombination and
for the full range of temperatures and densities examined here the
4471 \AA ~line has to be scaled by 6.6 $\cdot$ 10$^{-2}$
\citep[][]{Benjamin:99}.\linebreak

The age evolution tracks of \HeI$_{\rm 1.70 \mu m}$/Br10 are flat at
the youngest ages (for 0.4\Zsol ~and \Zsol), where the ratio is
saturated and the Str\"omgren spheres are both maximally filled
(Fig.~\ref{AgeEvol3}). With the incoming radiation field softening as
the cluster ages, we see the \HeI$_{\rm 1.70 \mu m}$/Br10 ratio drop
again in a similar manner to the mid-infrared line ratios. The age
evolution curves altogether have a very similar shape as the
mid-infrared curves, showing the W-R phase upturn and the rapid
decrease when the W-R stars disappear around 6/7 Myr. The ratio is
also mildly sensitive to density because of the presence of dust in
our models. When the density increases the dust competes more
effectively with the gas for FUV photons. The age evolution curves do
not show equally smooth behavior as those of the previously discussed
mid-infrared ratios, especially at the 2\Zsol ~W-R plateau between 3.5
and 6 Myr. This is most likely the result of numerical effects in the
radiative transfer of helium in the {\it Mappings} code. \linebreak

Another commonly used near-infrared diagnostic is the equivalent
width (EW) of one of the hydrogen lines as age estimator, usually the
bright ones like Pa$\beta$ in the J band or Br$\gamma$ in K. The
equivalent width measures the strength of the short-lived line
emission from the young, most massive stars relative to the continuum
emission from lower mass stars. Under the assumption of an
instantaneous burst of star formation, the EW drops with time
(EW(Br$\gamma$) is shown in Fig.~\ref{EW}).  The \Zsol ~and 2\Zsol
~curves drop monotonically, but in the low metallicity case the curve
shows a small jump between 3.5 and 4 Myr, correlated with the W-R
phase. With fewer heavy elements in the low metallicity W-R star
atmospheres, not a lot of ionizing photons are absorbed in the process
of line-blanketing and most are available for the ionization of the
surrounding matter. These ionizing photons boost the hydrogen line
fluxes. In the (super)solar metallicity case, where a lot of the
ionizing photons are used to ionize the W-R star atmospheres, this
upturn in the EW of Br$\gamma$ is not observed.  The decreasing value
of the EW with increasing density reflects the dusty nature of our
models in the same way as discussed for \HeI$_{\rm 1.70 \mu m}$/Br10.
For all but the most extreme case this results in a spread of less
than 33\% in the value of EW(Br$\gamma$). Only the models with an
ionization parameter $q$ of 8 $\cdot$ 10$^8$ in combination with a
density $n_H$ of 10$^6$ cm$^{-3}$ show very distinct values. The dust
becomes the dominant absorber of Lyman continuum photons in these most
extreme models.

\section{Comparison with existing mid-infrared models}

\begin{figure*}
\epsscale{1}
\plotone{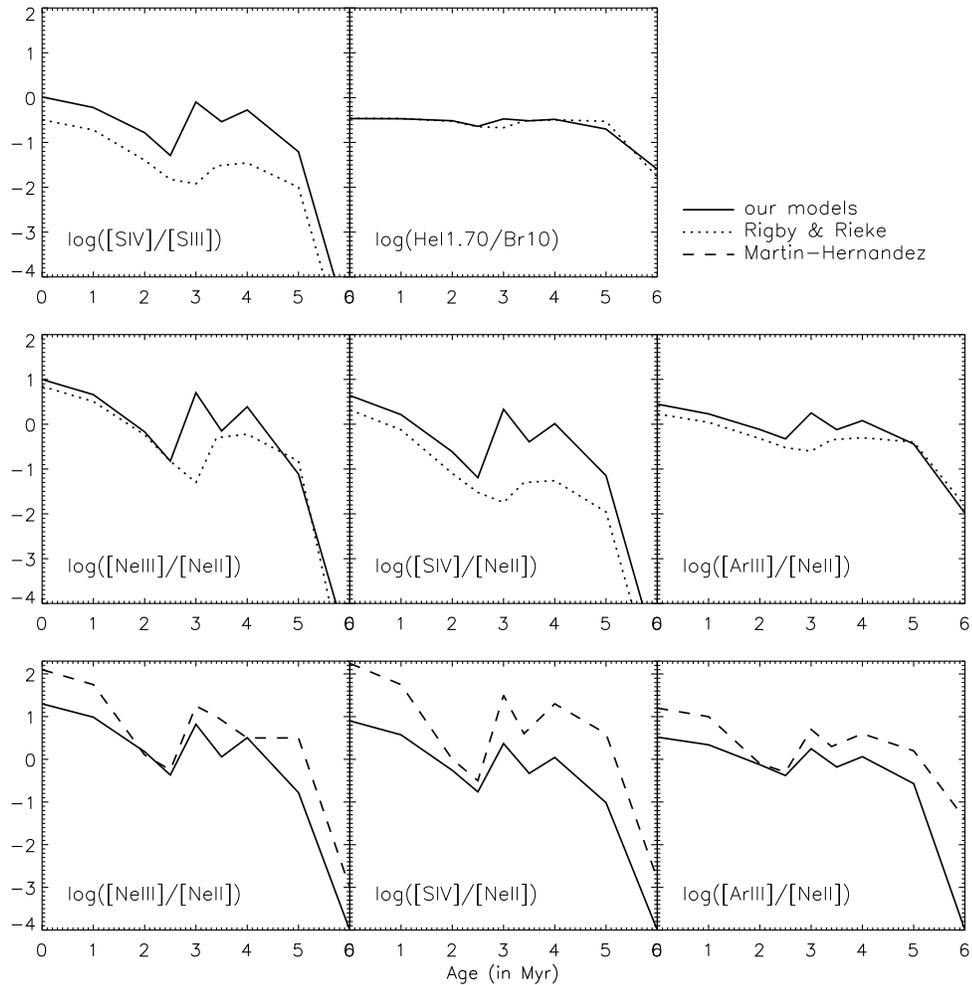}
\caption{Comparison of results for the age evolution of various
 emission line ratios of our models (solid lines) with results of RR04
 (dotted line in the top and middle row of panels) and with results of
 MH05 (dashed line in panels in the bottom row). The solid lines 
 (our model results) and the dotted curves (RR04) in the top and
 middle row of panels can be characterized by Z = \Zsol, $n_H$ = 300
 cm$^{-3}$ and $q$ = 1.37 $\cdot$ 10$^8$ cm s$^{-1}$. The solid lines
 and the dashed lines in the lowest row of panels show our model
 results and those of MH05 for Z = \Zsol,$q$ = 8.62 $\cdot$ 10$^9$ cm
 s$^{-1}$ and $n_H$ = 5 $\cdot$ 10$^4$ cm$^{-3}$.
\label{comp_ratio}}
\end{figure*}

Both \cite{Rigby:04} and \cite{Martin:05} (hereafter RR04 and MH05
respectively) have performed similar studies to this work, but
covering a more limited range of parameter space, assuming average
values for parameters such as density and ionization parameter. In
both works the age evolution of several line ratios was calculated in
a comparable way to that presented here, using {\it Starburst 99} in
combination with the photoionization code {\it Cloudy} \citep[][RR04
used {\it Cloudy} version 94.00, MH05 version 96.00-beta
4]{Ferland:01}. Both assume stellar populations with a Salpeter IMF
between 1 \Msol ~and various upper mass-cutoffs (RR04: 30, 40, 50, 60,
75 \& 100 \Msol ~and MH05: 30, 50 \& 100 \Msol), exploring a range of
sub-solar and solar metallicities. The models of RR04 are characterized
by an ionization parameter of log U = -2.3 ($q$ = 1.37 $\cdot$ 10$^8$
cm s$^{-1}$) and a density of 300 cm$^{-3}$. MH05 fix the density at 5
$\cdot$ 10$^4$ cm$^{-3}$. The ionization parameter for their object
(NGC 5253) is estimated to be at least log U = -0.5 ($q$ = 8.62
$\cdot$ 10$^9$ cm s$^{-1}$).

To accurately compare our models to those of RR04 and MH05, we
ran a set of models with identical parameters (solar metallicity, $q$
= 1.37 $\cdot$ 10$^8$, and $n_H$ = 300 cm$^{-3}$ for RR04, and solar
metallicity, $q$ = 8.62 $\cdot$ 10$^9$ cm s$^{-1}$ and $n_H$ 5 $\cdot$
10$^4$ cm$^{-3}$ for MH05). The resulting ratios of various emission
lines are plotted as a function of age in Fig.~\ref{comp_ratio}. As
judged from the age evolution of \HeI$_{\rm 1.70 \mu m}$/Br10 (upper
row in Fig.~\ref{comp_ratio}), the temperature evolution of the RR04
models is very similar to ours. The curves for this line ratio
essentially coincide. The mid-infrared line ratios of our models are
considerably different. Quantitatively the overall shape of the curves
is similar, with an initial drop and an upturn corresponding to the
W-R phase. However, the onset of the W-R phase happens earlier in our
models and the relative strength of the upturn is different; the W-R
stars do not seem to produce an equally hard radiation field in the
RR04 models as they do in ours. Since the discrepancy between the two
sets of models is largest during the W-R phase, we expect the modeling
of the W-R stellar atmospheres in {\it Starburst 99} to be (partly)
responsible for the differences. And although both studies use the
same atmospheric libraries, the treatment of W-R stars seem to have
changed between {\it v4.0} (used by RR04) and {\it v5.1} (used in our
work). Furthermore, there is a significant difference in the modeling
of the surrounding nebula. We include dust, while RR04 model a purely
gaseous nebula. The dust competes with the gas for ionizing photons,
which will significantly alter the ionization structure within the
nebula.  There are most probably computational differences between
{\it Mappings} and {\it Cloudy} as well, and differences in the
assumed value for the abundances, like sulphur over hydrogen (S/H)
(the abundance set is not given in RR04).

In the lower row of panels in Fig.~\ref{comp_ratio} we compare our
model results to those of MH05. The general shape of the age
evolution curves of MH05 shows a better agreement with our model
results than RR04. However, the agreement between the values of the
\NeIII/\NeII, \SIV/\NeII ~and \ArIII/\NeII ~line ratios produced by
our models and those of MH05 is worse, with those of MH05 often over
an order of magnitude larger. This difference most probably arises
from differences in the geometry applied. We model the gas and dust as
a thin slab (the thickness of the slab is much smaller than the
distance from the radiating source to the slab's inner radius) and
MH05 adopt a spherical model with a relatively thick shell of gas and
dust around the stellar cluster (inner radius is 0.6 pc, outer radius
is 0.8 pc). Because of this geometrical difference, the cloud in the
MH05 model sees an intenser UV field at the inner cloud radius for
identical cloud parameters. This results in large differences in
predicted mid-IR emission line ratios.

\section{Diagnostic diagrams}

\begin{figure*}
\begin{center}
\epsscale{1.2}
\plotone{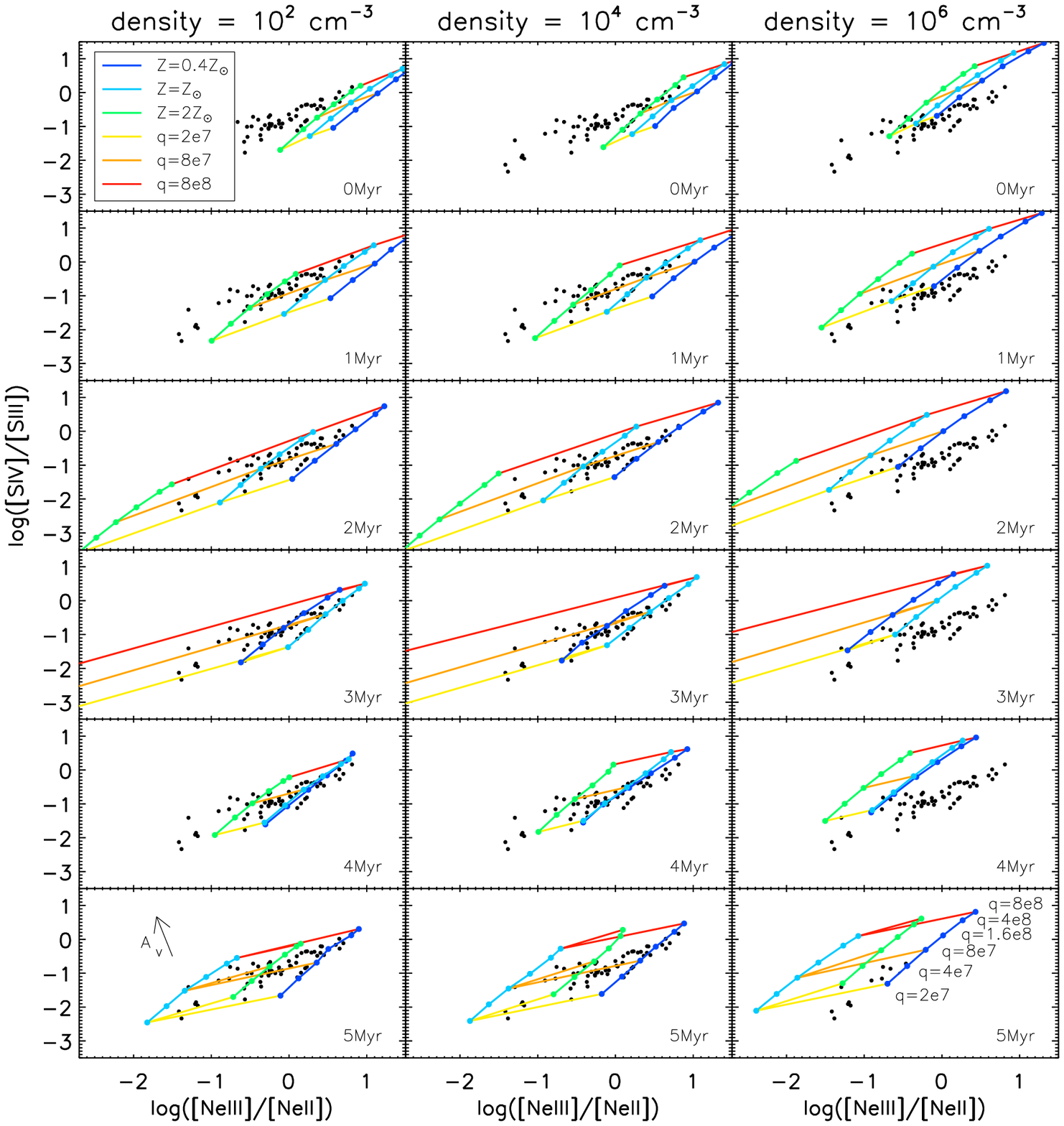}
\caption{\SIV/\SIII ~versus \NeIII/\NeII ~as a function of
metallicity, density and ionization parameter.  Stepping down from the
top to the bottom panels the plots show the age evolution of the
ratios in 1 Myr steps. The left column shows the lowest density case,
$n_H$ = 10$^2$ cm$^{-3}$, the middle column intermediate density,
$n_H$ = 10$^4$ cm$^{-3}$ and the right column shows the very dense
case, $n_H$ = 10$^6$ cm$^{-3}$. Within every individual panel, the
blue/green curves represent the different metallicities; dark blue
corresponds to Z = 0.4\Zsol, light blue to Z = \Zsol ~and green to Z = 2\Zsol. The
dots overplotted in the metallicity curves relate to the various
values for the ionization parameter $q$, with the lowest point
corresponding to $q$ = 2 $\cdot$ 10$^7$ cm s$^{-1}$ and the topmost
point to $q$ = 8 $\cdot$ 10$^8$ cm s$^{-1}$. All individual steps in
$q$ value are indicated in the bottom-right panel. The points for $q$
= 2 $\cdot$ 10$^7$ cm s$^{-1}$ are connected by a yellow line, the
points for $q$ = 8 $\cdot$ 10$^7$ cm s$^{-1}$ by an orange and those
for $q$ = 8 $\cdot$ 10$^8$ cm s$^{-1}$ by a red line. Note that the
metallicity curves start wrapping around each other $\ge$ 3 Myr. The
black dots are measurements of galactic \HII ~regions
\citep{Giveon:02}. Data points indicating upper limits are excluded
from this and the following plot to avoid confusion. The arrow in the
bottom-left panel indicates an A$_{\rm V}$ of 50. Line ratio values
can be found tabulated at www.ifa.hawaii.edu/$\tilde{\
}$kewley/Mappings/IRdiagnostics.
\label{S_Ne}}
\end{center}
\end{figure*}

To construct more powerful diagnostic tools, we combine various line
ratios into ratio--ratio plots. Figure~\ref{S_Ne} shows the age
evolution of \SIV/\SIII ~versus \NeIII/\NeII ~as a function of
metallicity, density and ionization parameter. 

Fig.~\ref{AgeEvol1} already showed that \SIV/\SIII ~and \NeIII/\NeII
~both measure the hardness of the radiation in a similar way and show
very comparable age evolution tracks. Therefore the curves run nicely
parallel and behave very orderly in the diagnostic diagram. Note
however that at ages $\ge$ 3 Myr the metallicity curves wrap around
each other, introducing a metallicity--spectral hardness
degeneracy. This effect is caused by the difference in sensitivity of
the ratios to radiation hardness. At this age the evolution curves in
Fig.~\ref{AgeEvol1} change rapidly and are substantially different for
each metallicity, causing the metallicity curves in the diagnostics
diagram in Fig.~\ref{S_Ne} to fold.

\begin{figure*}
\begin{center}
\epsscale{1.2}
\plotone{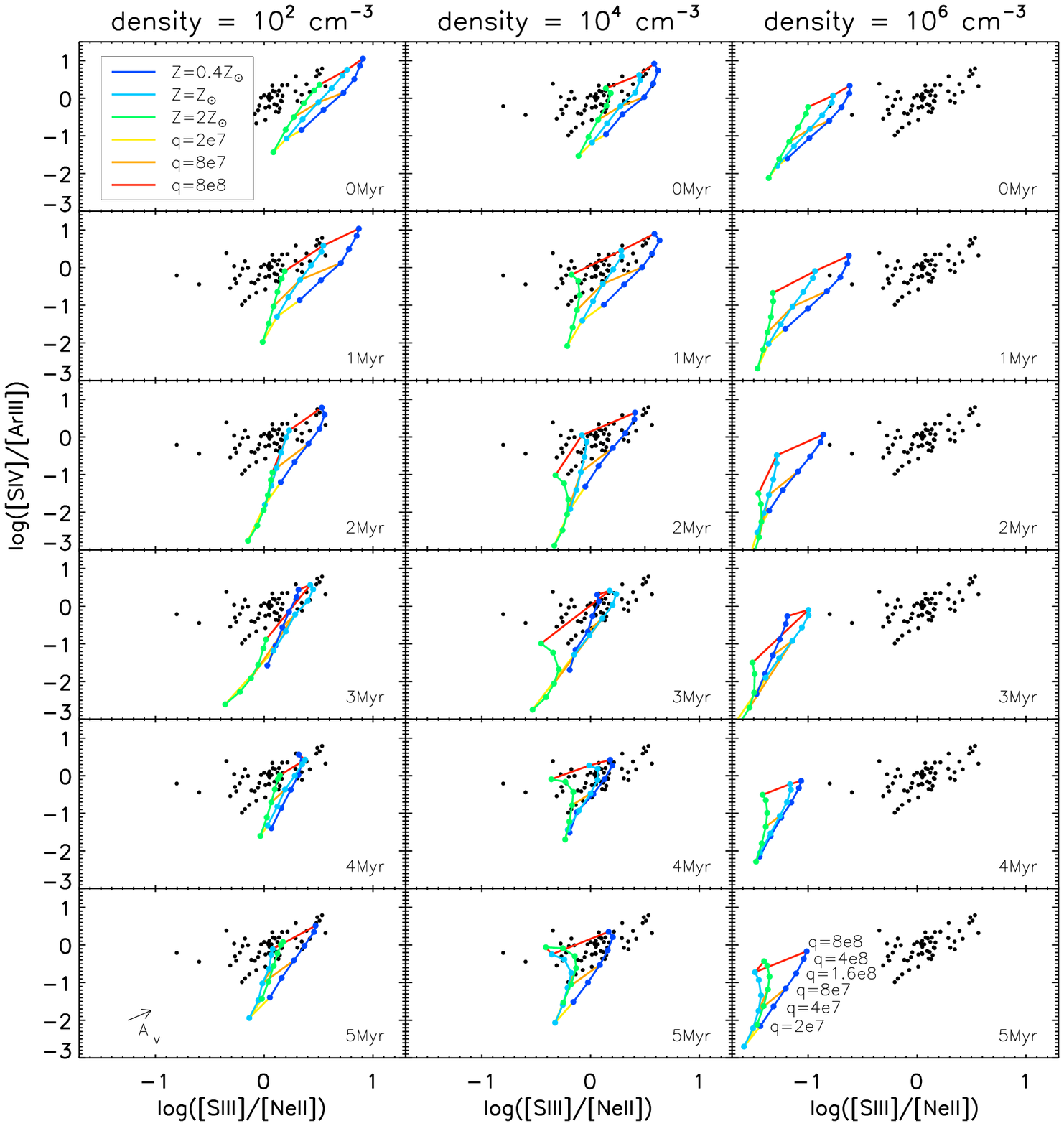}
\caption{As Fig.~\ref{S_Ne} for \SIV/\ArIII ~versus \SIII/\NeII. The
arrow in the bottom-left panel indicates an A$_{\rm V}$ of 100.
\label{SAr_SNe}}
\end{center}
\end{figure*}

The age evolution curves of \SIV/\ArIII ~and \SIII/\NeII ~differ
considerably in shape (Fig.~\ref{AgeEvol2}), causing the curves in the
ratio--ratio plot to look much more complex than the ones in the
diagram discussed in the previous paragraph. Different behavior from
the \NeIII/\NeII ~to \SIV/\SIII ~is expected, since here we plot a
hardness-sensitive ratio, \SIV/\ArIII, versus a ratio that is mainly
sensitive to density, \SIII/\NeII (Fig.~\ref{SAr_SNe}). The fact that \SIII/\NeII ~is
almost exclusively affected by density and not so much by radiation
hardness causes the relatively small spread in \SIII/\NeII ~values
within each individual panel. As in the \NeIII/\NeII ~to \SIV/\SIII
~diagram the metallicity curves wrap around each other at ages $\ge$ 3
Myr to form an interwoven knot of model points that are not easily
distinguished.

\section{Comparison with published data}

To test the model results, we compare the model output with
mid-infrared data of galactic and extragalactic \HII ~regions
published by \cite{Giveon:02}. This data set consists of a sample of
149 \HII ~regions; 112 galactic \HII ~regions, 14 \HII ~regions in the
Large Magellanic Cloud (LMC), 4 in the Small Magellanic Cloud (SMC)
and 19 in M33. From the ISO archive, the authors selected all \HII
~regions that are bright in one of the neon fine-structure lines and
for which both observations in low-resolution full spectrum mode
(Short Wavelength Spectrometer; SWS01) as well as in medium-resolution
line spectroscopy mode (SWS02) were available. The final spectra cover
the wavelength range of 2.4 -- 45 \um ~with a spectral resolution r =
400 (low resolution) - 2000 (medium-resolution). Since these
observations originate from numerous different projects, all with
their own distinctive research goals, the final sample is very
heterogeneous. It contains many regular star-forming regions as well
as very bright and/or compact \HII ~regions. This broad range of
conditions tests the ability of our model grid to reproduce a large
variety of different \HII ~region properties.

The data points of the \HII ~regions are plotted in the diagnostic
diagrams presented in the previous section (Figs.~\ref{S_Ne} and
\ref{SAr_SNe}). In Fig.~\ref{S_Ne} we see that the models and data
agree very well in general, occupying the same range in the
ratio--ratio values. The data points follow a distribution that is
somewhat tilted compared to the lines of equal ionization parameter,
most probably reflecting morphological differences between the
individual regions, resulting in variations in the ionization
parameter. The horizontal spread exhibits a range in metallicities and
densities. In general, the highest density models do not do as well
explaining the data as the low and intermediate density models
do. Furthermore, the very youngest age of 0 Myr can be ruled out for
most data points. On average the \HII ~data indicate an age for the
star-forming regions $\ge$ 1 Myr and an intermediate ionization
parameter (4 $\cdot$ 10$^7$ cm s$^{-1}$ $\le$ $q$ $\le$ 1.6 $\cdot$
10$^8$ cm s$^{-1}$).

Examining Fig.~\ref{SAr_SNe} shows that here again the data points and
models overlap nicely, although in this diagram there are a handful of
points displaying ratio--ratio values that are not covered by the
model grid, mainly at the high ionization end. It is quite clear that
a density of 10$^6$ cm$^{-3}$ is too high for all data points (but
one). On average the intermediate density model curves (10$^4$
cm$^{-3}$) cover a significant fraction of the data points, but all
densities ranging from 10$^2$ to 10$^5$ cm$^{-3}$ are capable of
reproducing the data. The mean ionization parameter indicated by
this diagram is higher than concluded from the previous diagram, for
most data a value of $q$ between 8 $\cdot$ 10$^7$ and 8 $\cdot$ 10$^8$
cm s$^{-1}$ is required.

\section{A test case from the ground: stellar clusters in the overlap region of the Antennae}

To address the diagnostics based on the spectral features that are
accessible from the ground only, we use infrared data of young stellar
clusters in the Antennae overlap region as a test case. We combine
near-infrared Infrared Spectrometer And Array Camera (ISAAC) at the
VLT \citep{Moorwood:98} with mid-infrared VLT Imager and Spectrometer
for mid Infrared \citep[VISIR, ][]{Lagage:04} spectra. 

The overlap region of the Antennae, the region of interaction between
the two merging spiral galaxies, is known from mid-infrared
observations to host the most active sites of star formation, some
severely reddened by the presence of massive amounts of dust. Based on
ISO observations \citet{Mirabel:1998} concluded that 15\% of the total
mid-infrared flux at 15 \um ~of the whole NGC 4038/4039 system
originates from a deeply embedded, off-nuclear star-forming region,
corresponding to a faint, very red cluster in the optical
\citep[source 80 in ][]{Whitmore:95}. The clusters under study are
this highly reddened cluster and an optically bright, blue cluster
complex, which is the counterpart of the second brightest infrared
source in the overlap region. The source IDs are 1a and 2 in
\cite{Snijders:06} for the reddened cluster and the cluster complex
respectively, corresponding to the bright near-infrared sources
ID$_{\rm WIRC}$ 157 and ID$_{\rm WIRC}$ 136 \citep{Brandl:05}.

\subsection{Observations, data reduction and results}

\begin{deluxetable*}{lccccc}
  \tabletypesize{\footnotesize} \tablecaption{Near- and mid-infrared data of star clusters in the Antennae overlap region \label{Ant}}
  \tablewidth{0pt} 
\tablehead{ \colhead{Species} &
            \colhead{Wavelength} &
            \multicolumn{2}{c}{Source 1a\tablenotemark{a}} &
            \multicolumn{2}{c}{Source 2\tablenotemark{b}} \\
	    &
	    \colhead{(\um)} &
            \colhead{measured} &
            \colhead{ext corr\tablenotemark{c}} &
            \colhead{measured} &
            \colhead{ext corr\tablenotemark{c}} 
}

\startdata

&&&&&\\
Br$\gamma$\tablenotemark{d}            & 2.166             & 1.4  $\pm$ 0.2     & 2.1  $\pm$ 0.2     & 1.5  $\pm$ 0.2     & 1.5  $\pm$ 0.2 \\
EW(Br$\gamma$)\tablenotemark{e}        & 2.166             & 3.6  $\pm$ 0.4     & 3.6  $\pm$ 0.4     & 2.8 $\pm$ 0.3 &  2.8 $\pm$ 0.3 \\
HeI$_{\rm 1.70 \mu m}$/Br10            & 1.700/1.737       & 0.309 $\pm$  0.046 & 0.315 $\pm$  0.047 & 0.299 $\pm$ 0.045 & 0.299 $\pm$ 0.045  \\
&&&&&\\
$[$Ar\,{\sc iii}$]$\tablenotemark{d}   & 8.99              & 3.9  $\pm$ 0.9    & 5.9  $\pm$ 1.4 & 4.2  $\pm$ 0.7      & 4.3  $\pm$ 0.7 \\
$[$S\,{\sc iv}$]$\tablenotemark{d}     & 10.51             & 7.6  $\pm$ 0.9    & 11.8 $\pm$ 1.4 & 10.3 $\pm$ 2.3      & 10.6 $\pm$ 2.4 \\   
$[$Ne\,{\sc ii}$]$\tablenotemark{d}    & 12.81             & 49.6 $\pm$ 6.2    & 59.8 $\pm$ 7.5 & 30.2 $\pm$ 4.5      & 30.6 $\pm$ 4.6 \\ 
$[$S\,{\sc iii}$]$\tablenotemark{d}    & 18.71             & 31.2 $\pm$ 3.8    & 38.4 $\pm$ 4.7 & 38.3 $\pm$ 3.9      & 38.8 $\pm$ 4.0 \\

  \enddata

\tablenotetext{a}{ID$_{\rm WIRC}$ 157 in \cite{Brandl:05}, cluster 80 in \cite{Whitmore:95}}
\tablenotetext{b}{ID$_{\rm WIRC}$ 136 in \cite{Brandl:05}, star-forming knot B in \cite{Whitmore:05}}
\tablenotetext{c}{A$_{\rm V}$ is 4.23 for source 1a and 0.28 for source 2}
\tablenotetext{d}{line fluxes in 10$^{-14}$ erg~s$^{-1}$~cm$^{-2}$}
\tablenotetext{e}{equivalent width in 10$^{-2}$ \um}

\end{deluxetable*}

The southern part of the overlap region of NGC4038/39 was observed
with VISIR. Low-resolution N band spectra of the two brightest sources
in the mid-infrared were already published in
\cite{Snijders:06}. Additionally, medium-resolution spectroscopy
around the \SIII ~line at 18.71 \um ~was obtained for the same two
sources. The total on-source integration time for these observations
was 80 minutes. The same setup was used as with the N band
observations, applying chopping and nodding on-slit with a 10\arcsec
~chopper throw. HD93813 was observed before and after the observation
for flux calibration \citep{Cohen:99}. The data were reduced in the
same way as the N band spectra and data were extracted in a 1\farcs27
~aperture ($\approx$ 133 pc at a distance of 21 Mpc\footnote[1]{Note
that the distance of the Antennae is under debate. Recently a lower
distance of 13.8 $\pm$ 1.7 Mpc was found \citep{Saviane:04}, which
would affect the values derived here.}, assuming a Hubble constant of
70 km~s$^{-1}$~Mpc$^{-1}$; at this distance 1\arcsec ~corresponds to
105 pc).

\begin{figure}
\epsscale{1}
\plotone{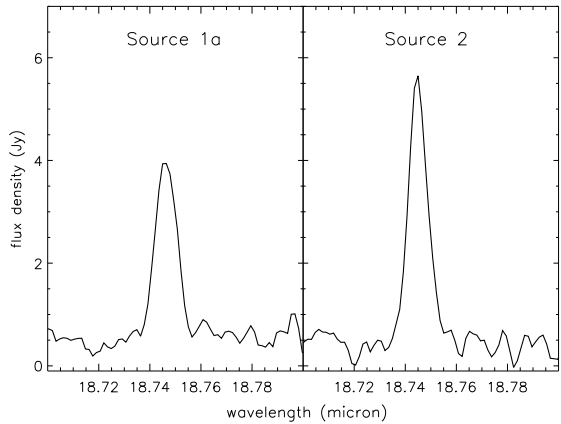}
\caption{Medium-resolution Q band spectroscopy of the 18.71 \um ~\SIII
~emission-line for {\it left:} source 1a, and {\it right:} source 2.
\label{Ant_Q}}
\end{figure}

Absolute flux calibration was obtained by normalizing the
standard star spectrum to VISIR narrow-band fluxes. The accuracy of
the flux calibration is 20\%. Figure~\ref{Ant_Q} shows the resulting
Q band spectra.

For our analysis we use the N and Q band emission lines, and
additionally the flux ratio of \HeI$_{\rm 1.70 \mu m}$/Br10 from the H
band and the EW(Br$\gamma$) from K$_{\rm s}$ band spectra, both
obtained with ISAAC (Snijders \& Van der Werf, in preparation). For the
near-infrared observations the on-source integration time was 20
minute in each spectroscopic setting. The spectra were extracted in
3\farcs5 apertures. This aperture is considerably larger than the
aperture applied on the mid-infrared data. We have done so to capture
all the hydrogen emission line flux that can be associated with the
clusters and to compensate for a worse seeing in the near- compared to
the mid-infrared.

The data were extinction corrected assuming a standard
\cite{Draine:89} extinction curve. We obtained estimates for the
visual extinction A$_{\rm V}$ from \cite{Mengel:05}. These values for
the extinction were derived from the ratio of the near-infrared
hydrogen recombination lines H$\alpha$ and Br$\gamma$. For source 1a
there was only one possible near-infrared counterpart, for which an
A$_{\rm V}$ of 4.23 was determined. For source 2 we selected all
sources within an 1\farcs4 radius, resulting in four possible
near-infrared counterparts. To correct source 2 for extinction we use
the average A$_{\rm V}$ value of these four sources, A$_{\rm V}$ =
0.28 (the spread in values for A$_{\rm V}$ was 0.16 --
0.42). Table~\ref{Ant} lists the mid-infrared emission line
fluxes and the EW(Br$\gamma$) and \HeI$_{\rm 1.70 \mu m}$/Br10 from
the near-infrared spectra.

\begin{figure}[b]
\begin{center}
\epsscale{1.2}
\plotone{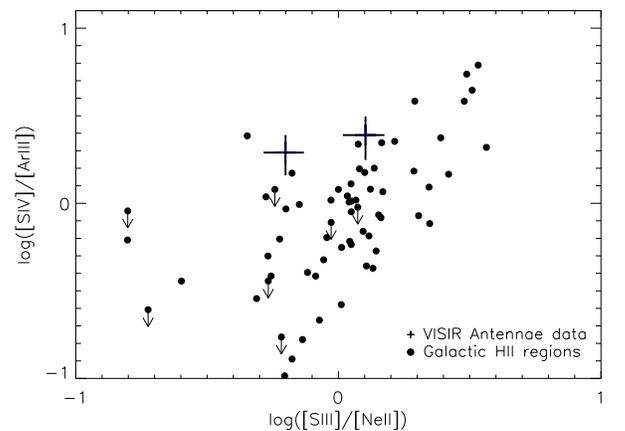}
\caption{Comparison of the \SIV/\ArIII ~and \SIII/\NeII ~ratio of the
Antennae clusters observed with VISIR (crosses, this work) with
(extra-)galactic \HII ~regions \citep[dots:][]{Giveon:02}. None of
the data is extinction corrected. The Antennae data points include
error bars. \label{data_VISIR}}
\end{center}
\end{figure}

The resulting mid-IR data points are compared with the sample of
\HII ~regions observed in our and other nearby galaxies (the data set
presented in Section 6). Fig.~\ref{data_VISIR} shows that the Antennae
clusters fall within the general distribution of \HII ~regions, though
right at the edge.

\begin{figure*}
\epsscale{1}
\plotone{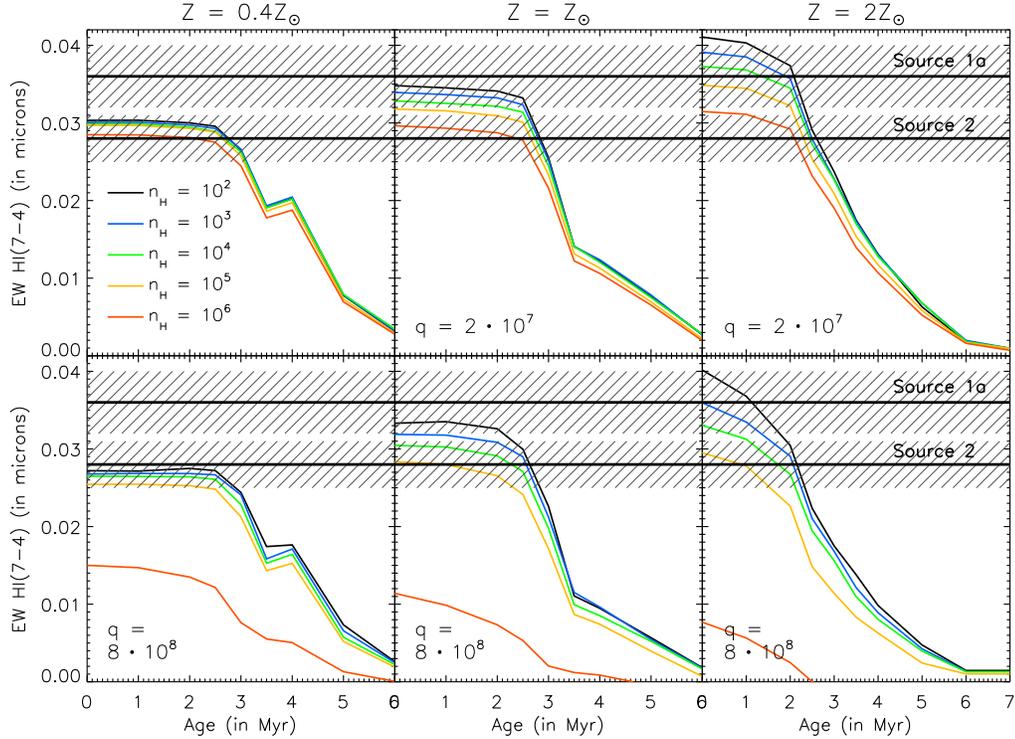}
\caption{This plot is identical to Fig.~\ref{EW} with the data points
for the Antennae clusters included. The fat black line indicates the value
of EW(Br$\gamma$) for both sources as measured from ISAAC
spectra. Error bars are indicated by the shaded areas. From this plot
we conclude that both clusters are young, source 1a $\le$ 2.5 Myr and
source 2 $\le$ 3Myr.
\label{EW_data}}
\end{figure*}

The cluster ages are constrained using the EW(Br$\gamma$) (see
Fig.~\ref{EW_data}). Source 1a has the highest value for
EW(Br$\gamma$) and its age must be 2.5 Myr or younger. Furthermore,
the metallicity of this source must be around solar or higher, since
the 0.4\Zsol ~cannot reproduce the observed value for EW(Br$\gamma$).
The age of source 2 falls in the range of 0 -- 3 Myr. For this source
it is not possible to discriminate between the different
metallicities. Taking these age constraints into account, it is clear
from the \SIV/\ArIII ~versus \SIII/\NeII ~diagnostic diagram
(Fig.~\ref{SAr_SNe_plusAnt}) that the highest density models, with a
$n_H$ of 10$^6$ cm$^{-3}$, disagree with the data points. In general
the medium density curves ($n_H$ = 10$^4$ cm$^{-3}$) come closest to
the data. Both clusters require a very high ionization parameter, the
data points occupy an area in the ratio--ratio plot that is
characterized by a value of $q$ $\ge$ 8 $\cdot$ 10$^8$ cm
s$^{-1}$. Further implications are discussed in Sections 8.2 and 8.3.

\newpage
\section{Discussion}

\subsection{Modeling infrared line emission of young star-forming regions}

The model grid presented in this paper covers a large range in
metallicities, gas densities and ionization parameters to mimic the
characteristics of all kinds of star-forming regions. Testing the
model output with galactic and extragalactic (UC)\HII ~regions shows
that the models succeed in reproducing the mid-infrared
characteristics of objects with a wide range of properties, from
diffuse \HII ~to UC\HII ~regions as well as starburst nuclei. There is
a very good agreement between the observed ratio--ratio values, like
\SIV/\SIII ~versus \NeIII/\NeII, and the range in values predicted by
the models.

\begin{figure*}
\epsscale{1.2}
\plotone{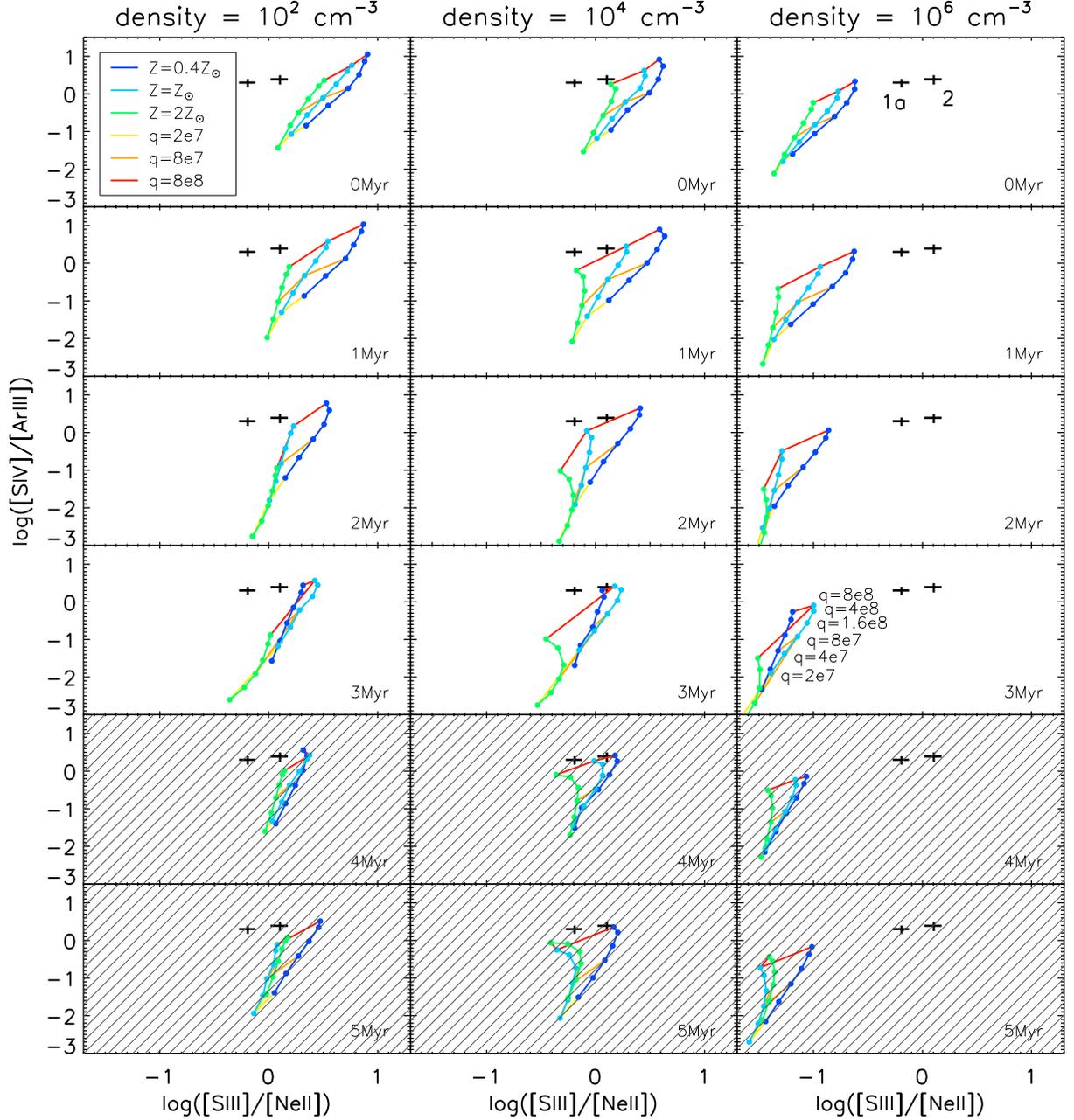}
\caption{This plot is identical to Fig.~\ref{SAr_SNe} with the
data points for the Antennae clusters included (crosses). The 4 Myr and
5 Myr panels are shaded, since we concluded from EW(Br$\gamma$) that
the clusters are 3 Myr or younger (source 2; source 1a is $\le$ 2.5
Myr). Error bars are given and the Antennae data points are extinction
corrected as described in the text.
\label{SAr_SNe_plusAnt}}
\end{figure*}

We do not encounter some of the problems reported in previous work on
the modeling of starburst galaxies. \cite{Kewley:01} finds in the
study of optical emission lines of starbursts that the FUV-field
produced by {\it Starburst 99 v4.0} is too hard to be able to
reproduce the observed line ratios (the number of photons in the 300
-- 1000 \AA ~range is too low compared to the highest energy photons
between 100 -- 300 \AA). As a possible solution the implementation of
line-blanketing in the atmospheres of high-mass stars was suggested,
which is indeed incorporated in the current version of {\it Starburst
99 (v5.1)}.  Nonetheless, the results of a more recent study of
mid-infrared observations of starburst galaxies show similar problems
\citep{Rigby:04}. Although these authors use a version of {\it
Starburst 99} with line-blanketing implemented, the observed
mid-infrared line ratios combined with \HeI$_{\rm 1.70 \mu m}$/Br10
can only be explained by models assuming an IMF with a relatively low
upper mass cutoff (40 -- 60 \Msol). In our work presented here there
is no need to assume an unusually low M$_{up}$. As discussed in
Section 4 we expect the fact that our models include dust, while the
Rigby \& Rieke models do not, plus changes in the modeling of W-R
stars in {\it Starburst 99} to be responsible for the differences
between our work and theirs. Furthermore, there may be computational
differences in the photoionization codes used; we use {\it Mappings}
and Rigby used {\it Cloudy}.

In spite of the improvements, several inconsistencies remain. The
average ionization parameter for Giveon's \HII ~regions derived from
the \SIV/\SIII ~versus \NeIII/\NeII ~plot, is lower than values
indicated by the diagram plotting \SIV/\ArIII ~versus \SIII/\NeII ~(4
$\cdot$ 10$^7$ cm s$^{-1}$ $\le$ $q$ $\le$ 1.6 $\cdot$ 10$^8$ cm
s$^{-1}$ for the first diagram and 8 $\cdot$ 10$^7$ cm s$^{-1}$ $\le$
$q$ $\le$ 8 $\cdot$ 10$^8$ cm s$^{-1}$ for the latter). This
discrepancy may originate from varying relative elemental abundances
and/or differences between the detailed shape of the stellar FUV
spectra in the models and in reality. Examining the data points in the
\SIV/\ArIII ~versus \SIII/\NeII ~diagram (Fig.~\ref{SAr_SNe}) suggests
that the values for \SIV/\ArIII ~predicted by the models are too low
and the strength of argon lines in the models is too high relative to
the sulphur emission lines. Difficulties in the modeling of sulphur
have been reported before (\cite{Martin:02,Verma:03}). However, in
these studies on ISO observations of galactic \HII ~regions and
starburst galaxies the sulphur abundance was found to be low compared
to neon and argon.

We modeled a very simple system, a single spherical stellar
population, formed in an instantaneous burst of star formation wrapped
in a thin shell of gas and dust. We assumed a Salpeter IMF between 0.1
and 100 \Msol ~for the stellar mass distribution of the
clusters. Although the shape of the IMF is under constant debate,
especially in extreme environments, such as in starbursts, \cite{Thornley:00}
showed that differences in metallicity and ionization parameter have a
much larger effect on mid-infrared line ratios than the shape of the
IMF. Given observational constraints, the assumption for the IMF upper
mass cutoff is not expected to introduce a large error either. From
the EW(Br$\gamma$) we know that both sources in the Antennae overlap
region are 3 Myr or younger. Furthermore, \cite{Kunze:96} derived an
effective temperature of 44,000 K from the \NeIII/\NeII ~ratio,
corresponding to the presence of a most massive star of 60 \Msol ~in
the present day mass function (PDMF). The variation of several
mid-infrared line ratios due to a different IMF upper mass cutoff
during the first 3 Myr is shown to be a factor of two or less
\citep[comparing M$_{up}$ of 50 \Msol ~and 100 \Msol ~in Fig.~3 in
][]{Rigby:04}. So the choice of upper mass cutoff in the range
constrained by the observed age and effective temperature has a
limited effect on the line ratios compared to varying density and
ionization parameter.

The assumption of an instantaneous burst of star-formation does have a
considerable impact on the mid-infrared emission lines. Because of the
constant production of massive stars, the SED of a stellar population
undergoing continuous star formation at a certain age is expected to
be harder than that of a passively evolving stellar cluster formed in
a single instantaneous burst of the same age. Thus under the
assumption of constant star formation the ISM properties do not need
to be as extreme (in density and ionization parameter) to reproduce
the observed line ratios. Another source of major uncertainty is the
modeling of stellar atmospheres. Since the mid-infrared emission line
ratios are very sensitive to the detailed shape of the FUV spectra,
the choice of atmospheric models have a large effect on the analysis
of the observations. Furthermore, we know from local \HII ~regions
that the morphology of star-forming regions is usually very complex
and nothing like a simple sphere. Since the regions discussed here are
large, with a radius of 40-45 pc, we expect significant clumpy
structure below our resolution limit. This issue will be addressed in
the next Section. Lastly we mention the presence of shocks as another
possible excitation mechanism. Indications for shocks being important
in the Antennae overlap region are found from extraordinarily bright
H$_2$ emission \citep{Haas:05}. Part of the line emission could be
shock excited, complicating the analysis.

\subsection{Dense star-forming regions in the Antennae}

Comparison of the model output with near- and mid-IR data of two
clusters in the Antennae overlap region allow us to constrain the
properties of the ISM surrounding the star clusters. Fitting
\SIV/\ArIII, \SIII/\NeII, EW(Br$\gamma$) and \HeI$_{\rm 1.70 \mu
m}$/Br10 simultaneously results in a good fit of source 2 by a 3 Myr
old stellar population with either solar or sub-solar metallicity, a
density of 10$^4$ cm$^{-3}$ and an ionization parameter of 8 $\cdot$
10$^8$ cm s$^{-1}$. Given that the metallicity of most star-forming
regions in the overlap region was found to be around solar from
optical observations \citep{Bastian:06}, we prefer the solution with
solar metallicity.

\begin{figure}
\epsscale{1}
\plotone{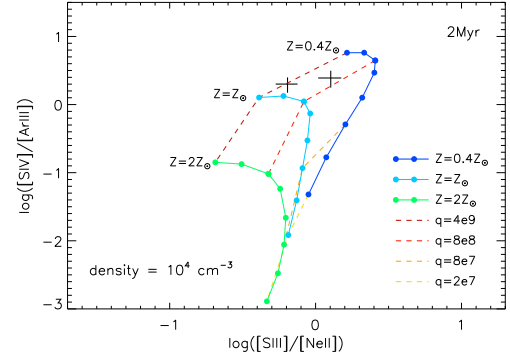}
\caption{A blow-up of one individual panel of
Fig.~\ref{SAr_SNe_plusAnt}. The metallicity curves have been extended
to higher values for the ionization parameter $q$. The solid curves
correspond to different metallicities, black to Z = 0.4\Zsol, dark grey to
Z = \Zsol, and light grey to Z = 2\Zsol ~(a color figures can be found in the
electronic edition).  The dots corresponds to $q$ = 2 $\cdot$ 10$^7$,
4 $\cdot$ 10$^7$, 8 $\cdot$ 10$^7$, 1.6 $\cdot$ 10$^8$, 4 $\cdot$
10$^8$, 8 $\cdot$ 10$^8$, 2 $\cdot$ 10$^9$ and 4 $\cdot$ 10$^9$ cm
s$^{-1}$. The grey dashed lines connect curves of equal $q$.
\label{SAr_SNe_highQ}}
\end{figure}

It is evident from Fig.~\ref{SAr_SNe_plusAnt} that, unlike source 2,
source 1a cannot be explained by a single component from the presented
model grid. One possible solution to reproduce the observed line
ratios is to extend the metallicity curves to higher values for the
ionization parameter. Under the assumption that the metallicity of
source 1a is around solar as well, the 2 Myr \Zsol ~curve for a
density of 10$^4$ cm$^{-3}$ looks most promising. To estimate by how
much $q$ must be increased we extended this curve by adding the data
points for a value for $q$ of 2 $\cdot$ 10$^9$ and 4 $\cdot$ 10$^9$ cm
s$^{-1}$. Figure~\ref{SAr_SNe_highQ} shows that the \SIV/\ArIII ~ratio
saturates for such high ionization parameter values. The line ratios
observed for source 1a can be fitted by a slightly super-solar
metallicity, 2 Myr stellar population, embedded in matter with a
density of 10$^4$ cm$^{-3}$ that can be characterized by an ionization
parameter of 4 $\cdot$ 10$^9$ cm s$^{-1}$. Compared to previous
estimates for $q$ this is a very high value \citep[see Fig.10 in][for
ionization parameters for a collection of local starbursts. The mean
value for logU is -2.3]{Rigby:04}. However, since the star-forming
regions in the Antennae are known to be very massive ($\sim$ one
million solar masses) and bright \citep{Mengel:01}, there is no reason
to rule out this option before further analysis. A measurement of the
\NeIII ~line flux would help to obtain more robust constraints on $q$
from our observations. Unfortunately it is impossible to obtain such
observations with the same spatial resolution as our ground-based
data. We note that a less extreme ionization parameter might be
achieved by the use of the ultra-compact \HII ~region models by
\cite{Dopita:06a}. This possibility will be explored in future work.

Since it is known from observations that star-forming regions are
usually made up of a clumpy, structured, multi-phase ISM, we consider
a second option to reproduce the mid-infrared characteristics of source
1a. We assumed a two-component medium by making linear combinations of model
output spectra (f$_1$ $\times$ spectrum$_1$ + f$_2$ $\times$
spectrum$_2$, with f$_1$ and f$_2$ varying from 0.1 to 0.9 in steps of
0.1 and f$_1$ + f$_2$ = 1; we only consider combinations with
identical metallicity). In this way one can obviously create a large
set of possible fits to the \SIV/\ArIII ~and \SIII/\NeII ~values of
source 1a. It is interesting though that, without exception, all these
possibilities require a high density. At least 70\% of the matter from
which these mid-infrared lines originate must have an average density
of $\ge$ 10$^5$ cm$^{-3}$. If we fit our model results to the mid- and
near-infrared lines simultaneously there is only one possible
combination that can explain the data of source 1a that meets the
assumption of solar metallicity: 90\% of the radiating ISM can be
characterized by a 0 Myr stellar population with a density of 10$^5$
cm$^{-3}$ and an ionization parameter of 8 $\cdot$ 10$^8$ cm s$^{-1}$
and the remaining 10\% by a 1 Myr solar metallicity population with a
10$^4$ cm$^{-3}$ and an ionization parameter of 8 $\cdot$ 10$^8$ cm
s$^{-1}$. In reality a two component medium is an oversimplification
as a model of an inhomogeneous star-forming region, but these results
still provide a good estimate of the range of average properties we
expect for these regions. 

Altogether combining the data and the models indicates that
source 1a is a very young stellar population packed in a giant cloud
of very dense gas and dust. During the first million years after the
birth of the stellar population, one might expect the cluster still to
be deeply buried in its natal cloud, reddened by far more obscuring
matter than the A$_{\rm V}$ of 4.2 derived from near-IR data. The
extinction is known to be patchy in this region, varying orders of
magnitudes on scales comparable to our spatial resolution
limit. Mid-IR narrow-band images revealed an object close to source 1a
suffering much more extinction \citep{Snijders:06}. Since this object
does not even have a counterpart detected in the near-IR, we concluded
that the A$_{\rm V}$ must be in the order of 70 or larger. However,
even if the extinction on source 1a is much larger than the A$_{\rm
V}$ of 4.2 corrected for here, our results would not change, since the
\SIV/\ArIII ~and \SIII/\NeII ~diagram is relatively insensitive to reddening
(see the vector corresponding to A$_{\rm V}$ = 100 in
Fig.~\ref{SAr_SNe}). 

\subsection{Evidence for a dense, clumpy ISM}

Comparing the average density found in this work with ISM densities in
the Antennae overlap region from the literature, we find that our
value is comparable to the density of molecular gas derived from
sub-mm $^{12}$CO and $^{13}$CO observations \citep{Zhu:03}. These
observations indicate a two phase molecular gas medium, one component
having a density $n_{H_2}$ of 1--8 $\cdot$ 10$^3$ cm$^{-3}$ and a
second denser component with $n_{H_2}$ $>$ 3 $\cdot$ 10$^4$ cm$^{-3}$,
both average values for the whole overlap region. \FeIII ~emission
lines in near-infrared spectra with comparable spatial resolution as
our data, indicate a somewhat lower ionized gas density than the value
we found, 10$^{3.5}$ -- 10$^4$ cm$^{-3}$ for source 1a \citep[under
the assumption that $n_H$ $\approx$ $n_e$. Source 2 is not addressed
in this paper, ][]{Gilbert:00}. In the same work a molecular gas
density of 10$^5$ cm$^{-3}$ is found from the analysis of H$_2$ line
emission.

Similar densities are locally observed only on very small scales in
UC\HII ~regions. Typically these star-forming clouds have a molecular
gas density $\ge$ 10$^5$ cm$^{-3}$ within a radius $\le$ 0.5 pc
\citep{Churchwell:02}. With comparable average gas densities, but
radii almost two orders of magnitude larger \citep[the radius of
source 1a and 2 is approximately 45 pc and 40 pc
respectively][]{Snijders:06}, the star-forming regions in the
Antennae bear more resemblance to extreme star-forming regions in
ultra luminous infrared galaxies (ULIGs). A large fraction of the
molecular gas in these sources is found to have densities larger than
10$^4$ cm$^{-3}$ \citep{Solomon:92}. In a large sample of ULIGs
\cite{Solomon:97} find somewhat lower molecular gas densities of
10$^3$ -- 10$^4$ cm$^{-3}$ for regions of 100 -- 500 pc radius. In a
subsequent paper, similar densities (2 $\cdot$ 10$^3$ -- 2 $\cdot$
10$^4$ cm$^{-3}$) are found for starburst regions in Arp193, Mrk273
and Arp220 for regions with 68 -- 150 pc radii. Gas masses for these
regions vary from 0.6 to 1.1 $\cdot$ 10$^9$ \Msol ~\citep{Downes:98}.

If we assume a homogeneous distribution of the gas in spherical clouds
the gas mass for source 1a is $\sim$ 1.2 $\cdot$ 10$^8$ \Msol ~(gas
density $n_H$ of 10$^4$ cm$^{-3}$, taking the presence of helium into
account, and a radius of 45 pc) and $\sim$ 8.6 $\cdot$ 10$^7$ \Msol
~for source 2 (gas density $n_H$ of 10$^4$ cm$^{-3}$ and a radius of
40 pc). However, the assumption of a homogeneous medium is
inconsistent with reality, as shown by the following argument. If we
combine the results for the ionization parameter and the density
derived from our models with the Lyman continuum emission calculated
from the extinction corrected Br$\gamma$ flux (Table~\ref{Ant}), we
can get an estimate for the source size using Eq.~\ref{eq:q}. Under
the assumption of case B recombination Q$_{\rm Lyc}$ is approximately
9.3 $\pm$ 0.9 $\cdot$ 10$^{52}$ s$^{-1}$ and 6.7 $\pm$ 0.7 $\cdot$
10$^{52}$ s$^{-1}$ for source 1a and 2 respectively (this is in good
agreement with determinations of Q$_{\rm Lyc}$ for source 1a by
Gilbert et al. 2000, from near-infrared emission hydrogen lines and
Hummel \& Van der Hulst 1986, from thermal radio continuum flux, both
find 1 $\cdot$ 10$^{53}$ s$^{-1}$). If we focus on source 1a, and
evaluate Eq.~\ref{eq:q} with the appropriate value for Q$_{\rm Lyc}$,
a density $n_H$ of 10$^4$ cm$^{-3}$ and a value for $q$ of 4 $\cdot$
10$^9$ cm s$^{-1}$, the radius of the emitting source should be $\sim$
4.4 parsec (using the values for the two-component fit, $n_H$ = 10$^5$
cm$^{-3}$ and $q$ = 8 $\cdot$ 10$^8$ cm s$^{-1}$, results in a radius
of 3.1 pc). This value is an order of magnitude smaller than the
source radius measured from mid-infrared images, namely a half-light
radius of 45 parsec.

This discrepancy implies the need for a more realistic, physical model
than a single stellar population surrounded by a single-phase gas
cloud. We adopt the geometry of a collection of $N$ distinct \HII
~regions (star-forming clumps) embedded in a giant molecular cloud,
very similar to the proposed configuration in Fig.~7 of
\cite{Forster:01} for the nearby starburst M82. Observational support
for this model can be found in mid-infrared images of source
1a. Although most of the clumpy substructure would be
indistinguishable because of the spatial resolution of our
observations, the general idea agrees with the finding of multiple
ionizing sources embedded in an envelope of continuum and PAH emission
\citep{Snijders:06}. Star clusters in cluster complexes are known to
form following a certain mass distribution \citep[][ and references
therein]{Fall:06, Dopita:06b}, so ideally one should perform Monte
Carlo simulations of various different stellar populations buried in a
clumpy ISM. However, for simplicity we assume $N$ identical stellar
populations. The Lyman continuum originating from each of these
individual clusters is Q$_{\rm Lyc}$/$N$. Combining the equation for
the ionization parameter $q$ for each individual region with the
ionization balance for a Str\"omgren sphere (Eqs.~\ref{eq:q2} and
\ref{eq:Stromgren}), results in an indication of the size and number
of individual \HII ~regions.

\begin{equation}
\label{eq:q2}
q = \frac{\frac{Q_{\rm Lyc}}{N}}{4\pi R^2n_{\rm ion}} 
\end{equation}

\begin{equation}
\label{eq:Stromgren}
\frac{Q_{\rm Lyc}}{N} = \frac{4}{3}{\pi R_{\rm S}^3n_{\rm ion}^2{\alpha}_{\rm B}} 
\end{equation}

$\alpha_{\rm B}$ is the recombination rate coefficient, 2.6 $\cdot$
10$^{-13}$ cm$^3$ s$^{-1}$ at T$_{\rm eff}$ = 10,000 K. $R$ is the
distance between the radiating source to the inner cloud boundary and
$R_{\rm S}$ is the Str\"omgren radius. For typical \HII ~regions most
\HII ~radiation originates from an ionized shell of gas around the
Str\"omgren radius, so $R$ $\approx$ $R_{\rm S}$. Solving this set of
equations with identical values as used above results in a typical
radius of 1.35 parsec and a number $N$ of $\sim$ 9 clumps for source
1a. The Lyman continuum implies that within each of these clumps
several hundreds of O-type stars are present \citep[Q$_{\rm Lyc}$ =
9.6 $\cdot$ 10$^{51}$ s$^{-1}$ corresponds to the ionizing flux from
$\sim$ 225 O3 stars, using the calibration of O stars
from][]{Martins:05}. The total ionized gas mass from all clumps added
up is 2.8 $\cdot$ 10$^5$ \Msol, approximately 3.5 orders of magnitude
less than under the assumption of a homogeneous cloud of gas.

A nearby example of a system comparable to the model proposed here, a
giant star-forming complex encompassing a handful of bright \HII
~regions, can be found in M82 \citep[at 3.6 Mpc distance, see Fig.1
in][]{Smith:06}. Region A, an elongated cluster complex of ~$\sim$ 50
-- 100 pc, shows rich substructure with several tens of star-forming
clumps.  The characteristics of one of these massive stellar clusters,
M82-A1, is very similar to the properties inferred above for the
star-forming clumps: the compact \HII ~region has a radius of 4.5
$\pm$ 0.5 pc and a density $n_e$ of 1800 cm$^{-3}$. The central
stellar population has an estimated age of 6.4 Myr and the total
stellar mass is around a million solar masses. At the distance of the
Antennae most information on the rich structure observed in M82's
cluster complex A would be lost due to spatial resolution effects. It
is very plausible that at a younger age the cluster complex would look
very similar to one of the bright SSCs in the Antennae overlap
region.

Further constraints on the stellar content of the star-forming clumps
in our model can be determined by estimating the total stellar mass
for source 1a by comparing the extinction corrected Br$\gamma$ flux
with the {\it Starburst 99} output for a Salpeter IMF between 0.1 and
100 \Msol. Depending on age, which we constrained to be 0 -- 2.5 Myr
for source 1a, the total stellar mass is between 3.2 $\pm$ 0.3 $\cdot$
10$^6$ \Msol ~(0 Myr) and 4.2 $\pm$ 0.4 $\cdot$ 10$^6$ \Msol ~(2.5
Myr; at the lower distance as found by \cite{Saviane:04} the mass
would be lower by a factor of $\sim$ 2.5). This is in good agreement
with the value of 3 million solar masses found by \cite{Mengel:01},
but considerably lower than the 16 $\cdot$ 10$^6$ \Msol ~by
\cite{Gilbert:00}. Note however that the studies of Gilbert and Mengel
both use a lower mass cutoff of 1 \Msol, and Gilbert adopts an age of
4 Myr. With a total stellar mass of 3 -- 4 million solar masses, each
individual star-forming clump would host 3 -- 4 $\cdot$ 10$^5$ \Msol
~of stars, resulting in a stellar density of 3 -- 4 $\cdot$ 10$^4$
\Msol ~pc$^{-3}$, which is equivalent to a density $n_H$ of 0.9 - 1.2
$\cdot$ 10$^6$ cm$^{-3}$. This implies very high star-formation
efficiencies if we assume these stellar populations to be formed from
typical dense molecular clouds (cloud core densities $n_H$ $\approx$
10$^6$ -- 10 $^7$ cm$^{-3}$). Studies of the most massive galactic
star-forming regions and of super star clusters (SSC) in various
starburst galaxies found comparable equivalent densities, as
summarized in Fig.~1 by \cite{Tan:06}. In this Figure source 1a would
end up between the Arches cluster in the Milky Way and the central SSC
in NGC5253.

Although our findings for the average stellar mass densities are
similar to results for other massive star-forming regions, there are
scenarios that require less extreme mass densities. For a stellar mass
distribution with relatively little low-mass stars compared to
'Salpeter' (e.g. a Miller-Scalo, Kroupa, Chabrier, or top-heavy IMF),
the mass within the 1.35 pc radius would be less. Indications for the
existence of an IMF dominated by massive stars is found in several
massive star-forming regions, in the center of our galaxy
\citep{Nayakshin:05} as well as in other starburst galaxies
\citep{Doane:93,Rieke:93}. Such a deviant IMF could be the result of
stellar winds and supernova explosions of the high-mass stars shutting
off the slower formation process of low-mass stars by expelling the
available matter from the star-forming region. Another option is mass
segregation within the stellar cluster. The high-mass stars in
clusters are often observed to be more centrally concentrated than the
low-mass stars. This is possibly caused by rapid inward migration of
massive stars \citep{Freitag:06,Portegies:99}. The mid-infrared
emission lines we observe, originate predominantly from the highest
mass stars, and the cluster core, where most of these massive stars
reside after mass segregation, would be deficit in low-mass
stars. This mechanism would not affect the total cluster mass, but
would lower the derived density in the cluster core.

Another plausible model for the morphology of a cluster complex would
be several star-forming clumps embedded in a large cloud of diffuse,
ionized gas. However, from our mid-infrared data we do not see
evidence for a significant fraction of low density ionized gas. Gas
with a density of 10$^2$ cm$^{-3}$ would boost the fluxes of the
emission lines with the lowest critical densities (e.g. \SIII ~at
33.48 \um). Consequently, the presence of a large amount of low
density ionized gas would leave its fingerprints on the observed line
ratios, which would have been recovered as possible solutions in the
two-component medium fits. The only possibility would be a component
of low density gas characterized by a very high ionization parameter
($q$ $\ge$ 5 -- 10 $\cdot$ 10$^9$ cm s$^{-1}$). As we have seen in
Fig.~\ref{SAr_SNe_highQ} the \SIV/\ArIII ~ratio saturates at high
values of $q$, causing a bend in the metallicity curves towards lower
\SIII/\NeII ~values. In this way it would be possible to have a
significant contribution from low density gas to the observed line
ratios. We consider this option highly unlikely though. A combination
of low density and such high ionization parameter would not arise,
since $q$ is inversely proportional to $R^2$ and we expect the diffuse
gas to be at a relatively high distance from the ionizing sources on
average, which would suppress $q$ (Eq.~\ref{eq:q}). Furthermore, the
typical ionization parameters for star-forming regions in starburst
galaxies derived from observations are lower \citep[Fig.~10
in][]{Rigby:04}, as well as the values for $q$ predicted by models
\citep{Dopita:06a}.

\section{Conclusions}

The recent advent of the first generation of mature mid-infrared
instruments on ground-based telescopes was one of the main motivations
for this work. With mid-infrared observations being performed
routinely on various 8-meter class telescopes nowadays, rich new
data sets rapidly become available. A lot of work has been done based
on space-based mid-infrared observations over the last
decades. Because of the characteristics of the earth's atmosphere,
ground-based work is limited to the spectral features observable from
the atmospheric windows of reasonably good transmission. So, in this
work we focus on the features that are available from ground-based
observations.

We have used the stellar population synthesis model {\it Starburst 99}
combined with the photoionization code {\it Mappings} to simulate
mid-infrared emission lines of star-forming regions in starburst
galaxies. We have shown that the models succeed in reproducing the
observed line ratios of \HII ~and starburst regions with a wide
variety of properties, ranging from diffuse to very dense systems.

Applying the model results in an analysis of the mid-infrared data of
SSCs in the Antennae overlap region, leads to the conclusion that the
average ionized gas density in these star-forming region is very
high. Source 2, the mid-infrared counterpart of a blue cluster
complex, can best be modeled by a 3 Myr old stellar population
embedded in dense matter, $n_H$ = 10$^4$ cm$^{-3}$, that can be
characterized by an ionization parameter of 8 $\cdot$ 10$^8$ cm
s$^{-1}$. The observed line ratios for source 1a, corresponding to the
highly reddened cluster WS95-80 in \cite{Whitmore:95}, can only be
reproduced by assuming an extremely high ionization parameter $q$ = 4
$\cdot$ 10$^9$ cm s$^{-1}$ in combination with a 2 Myr stellar
population and 10$^4$ cm$^{-3}$ density. We also considered a
two-phase ISM by fitting linear combinations of model output. This
resulted in a younger stellar population of 0 -- 1 Myr, surrounded by
10$^5$ cm$^{-3}$ gas characterized by $q$ = 8 $\cdot$ 10$^8$ cm
s$^{-1}$.

Detailed comparison of the mid-infrared emission lines with the images
shows us that the ISM in these two star-forming regions is as expected
far from homogeneous. As a possible geometrical model we propose
several individual star-forming clumps embedded in a giant molecular
cloud. This model is supported by the morphology of the source in the
mid-infrared images, showing two ionizing sources within a large
enveloping cloud of continuum and dust emission. The typical number of
\HII ~regions within the molecular cloud of $\sim$ 40 pc radius is 9,
each with a radius in the order of 1.35 pc (source 1a). Such
morphology is observed in nearby starbursts as well. Each of these
clumps would contain $\sim$ 225 O3 stars and have a gas density of 320
\Msol/pc$^3$ ($n_H$ = 10$^4$ cm$^{-3}$). \linebreak

\acknowledgments

We thank the Paranal Observatory Team for their support. Furthermore,
we thank Brent Groves for advice and fruitful discussions that
improved the paper. Part of this work was funded by the Leids
Kerkhoven-Bosscha Fonds, we thank them for their support.

\end{document}